\documentclass[10pt,3p,times,authoryear]{elsarticle}

\usepackage{amssymb}
\usepackage{amsmath}
\usepackage{hyperref}
\usepackage{orcidlink}
\usepackage{algorithm} 
\usepackage{algpseudocode}
\usepackage[version=4]{mhchem}

\journal{arXiv}

\begin{document}

\begin{frontmatter}

\title{Dissolution-induced mode transition of crack propagation in carbonate rocks: From brittle fracture to diffuse damage}

\author[HKU,DUKE]{Fanyu Wu \orcidlink{0000-0003-3197-0600}}
\ead{fywu@connect.hku.hk}
\author[HKU,Utrecht]{Chong Liu \orcidlink{0000-0002-0091-0130}}
\ead{chongliu@connect.hku.hk}
\author[ZJU,HKU]{Mingze Hu}
\ead{mingze0827@gmail.com}
\author[DUKE]{Manolis Veveakis \orcidlink{0000-0002-4911-6026}}
\ead{manolis.veveakis@duke.edu}
\author[ZJU,HKU]{Man-man Hu\corref{cor1} \orcidlink{0000-0001-5091-0289}}
\cortext[cor1]{Corresponding author.}
\ead{mmhu@zju.edu.cn}

\affiliation[HKU]{organization={Department of Civil Engineering},
                    addressline={The University of Hong Kong}, 
                    city={Hong Kong},
                    country={China}}

\affiliation[ZJU]{organization={Department of Civil Engineering},
                    addressline={Zhejiang University}, 
                    city={Hangzhou},
                    country={China}}
\affiliation[DUKE]{organization={Civil and Environmental Engineering},
                    addressline={Duke University}, 
                    city={Durham},
                    state={NC},
                    country={USA}}
\affiliation[Utrecht]{organization={Department of Earth Sciences},
                    addressline={Utrecht University}, 
                    city={Utrecht},
                    country={The Netherlands}}

\begin{abstract}
In chemically reactive environments, the mechanical integrity of geomaterials is fundamentally compromised by solid matrix dissolution. In this study, we propose a fully coupled chemo-mechanical phase-field framework to capture the dynamic interplay between mineral dissolution and fracture propagation. A key feature of the proposed model is the dynamic coupling of local mass removal to the fracture length scale, while also incorporating the damage-accelerated reaction-diffusion processes. Our results capture the development of an enlarged fracture process zone driven by chemical mass removal. This chemically induced widening blunts the sharp crack tip, alleviating the near-tip stress concentrations and causing a pronounced degradation in material stiffness before failure. Furthermore, we reveal a distinct ductilization effect, characterized by a more gradual accumulation of damage and a delayed onset of macroscopic failure. We show that the transition between brittle and ductile failure modes is dictated by the competing timescales of chemical degradation and mechanical deformation. Highly acidic environments enhance matrix dissolution and promote ductile fracture, whereas rapid mechanical loading limits chemical interaction and preserves brittle failure mode.
\end{abstract}

\begin{highlights}
\item We propose a novel chemo-mechanical phase-field fracture model for rocks subject to acidic environment.
\item The model captures a chemically induced length-scale of fracture process zone around the crack-tip.
\item A brittle-to-ductile transition in the fracture mode is identified due to time-dependent chemical weathering.
\item Competition between chemical mass removal rate and the counterpart of mechanical loading is investigated.
\end{highlights}


\end{frontmatter}


\section{Introduction}
\label{sec1}

Controlling the integrity of geomaterial is paramount for the long-term safety of critical subsurface energy applications, including enhanced geothermal systems (EGS), geological carbon storage (GCS), and nuclear waste disposal~\citep{rutqvist_numerical_2005,vafaie_chemo-hydro-mechanical_2023,horne_enhanced_2025}. These operations involve either the injection of reactive fluids, such as through hydraulic fracturing and acidizing treatments, or the long-term exposure of geomaterials to chemically aggressive environments, for example the acidic brine formed during CO$_2$ sequestration~\citep{economides_reservoir_2000,ilgen_coupled_2019}. In such reactive systems, the chemical processes fundamentally alter the mechanical properties of the solid matrix through mineral dissolution coupled with damage evolution~\citep{castellanza_oedometric_2004,ciantia_weathering_2013,stefanou_chemically_2014,ciantia_effects_2015}. During the progressive degradation of rock integrity, chemical weakening occurs originating from mass removal at both inter-granular and intra-granular scales, promoting internal deterioration and propagation of pre-existing cracks~\citep{buscarnera_chemo-mechanics_2016,tang_effect_2025,wu_onset_2025}. Localized damage zones and tensile fractures provide effective pathways for leakage of fluids, leading to undesirable regional or system-spanning failure~\citep{zoback_reservoir_2007,chen2023shear}. These large-scale instabilities are rooted in the mineral mass removal process taking place at available reactive sites down to nanometer scales, linking to the progress of volumetric micro-cracking ~\citep{wu_influence_2025}.

Thereby, dissolution-driven chemical softening of a dissolvable geomaterial is inherently a cross-scale phenomenon; understanding the dynamic interplay between its bulk mechanical response and the prescribed reactive environment is crucial~\citep{hueckel_feedback_2009}. Meanwhile, growing attention has been paid to the reaction diffusion process where the prescribed environment surrounding the solid mass acts as a reactive source that gradually diffuses into the stressed matrix as the chemical reaction progresses. The nonlinearity of crack propagation in a rock matrix over time is attributed to both the progressive mineral dissolution (defined by the chemical reaction rate) and the diffusive transport (controlled by the diffusion constant)~\citep{chen_chemically_2025,chen_fluid_2026}. Nevertheless, capturing and modeling this complex feedback in the context of subcritical cracking, where mineral mass removal and solid matrix deterioration are tightly coupled within the process zone around the crack-tip, remains a significant challenge. 

Numerical fracture modeling falls into two general classes. Discrete (sharp‑crack) approaches, such as cohesive‑zone models and extended finite element method, explicitly represent discontinuities but typically require ad‑hoc initiation or branching criteria and prior knowledge of crack paths~\citep{dugdale_yielding_1960,belytschko_elastic_1999,moes_finite_1999,liu2018}. Continuum (smeared) formulations represent cracks as diffuse damage fields, among which, the phase‑field method is especially attractive because a single continuous order parameter captures initiation, propagation, branching and merging without explicit crack tracking, offering numerical robustness and ease of implementation~\citep{francfort_revisiting_1998,bourdin_numerical_2000,borden_phase-field_2016}. Computational formulations based on the phase-field method have hence been developed to address cracking in a variety of solids, including metallic materials~\citep{martinez-paneda_phase_2018,kristensen_phase_2020,cui_phase_2021}, hydrogels~\citep{liu2026role}, cemented granular media~\citep{guevel_darcy_2023,wu_onset_2025}, composites~\citep{li_meso-scale_2024,s_fracture_2025,mishra_fracture_2026}, as well as geomaterials~\citep{chukwudozie_variational_2019,fei_phase-field_2023,liu_generalized_2026}.

The phase-field method has been extensively adopted for simulating multiphysics geomechanical scenarios, with demonstrated capability in modeling fracture phenomena coupled with thermal processes or the hydraulic field that is operating at comparable time-scales~\citep{miehe_phase_2016-1,ruan_thermo-mechanical_2023,wu_crack_2023}. 
Attempts on extending the phase-field method to chemo-mechanical problems were subsequently made, taking into account the reactive transport processes ~\citep{miehe_phase_2016}, which are common features of a wide range of fluid-driven applications in the subsurface. By far, two primary categories of modeling strategy exist in the literature that implement the coupling of chemical processes into phase-field fracture models -- considering geomaterials reactive to the prescribed chemical environment. 

The first category adopts a one-way coupling scheme that emphasizes the influence of stress state on the reactive transport occurring in the pore space while neglecting the effect of chemical process on the matrix stress evolution. For instance, \citet{wu_phase-field_2016} investigated the effect of cement cracking on the Fickian diffusion of chloride ions by introducing an anisotropic diffusivity tensor as a function of the phase-field parameter. Their results show that the diffusivity of ions parallel to the newly created crack plane can be notably increased. 
The second category addresses a two-way coupling between the mechanics of degrading solid matrix and the chemistry of the interstitial fluid, which is gaining increasing popularity for fracture modeling. 
This is enabled by introducing a chemical damage variable that separates from the mechanical phase-field variable. 
The chemical damage, often defined as a function of the change in hydraulic properties resulting from mineral dissolution, then acts in concert with the mechanical damage to degrade the elastic strain energy. Concurrently, the mechanical field influences the chemical process via the phase-field variable or the computed strain. 

A few notable contributions on shaping the two-way coupling phase field method for studying fracturing in geomaterials exposed to chemically reactive environment are addressed as follows, placing a specific emphasis on dissolution induced chemical degradation.  \citet{schuler_chemo-mechanical_2020} established a link from calcite dissolution to porosity evolution and chemical damage. The formulation allows hydrostatic stress gradients to drive solute flux while the phase-field and porosity continuously update the effective diffusivity. \citet{guo_reactive-transport_2024} extended this strategy to address poroelastic multi-mineral sandstone acidizing, where acid-species transport and mineral mass balances modify hydraulic properties and chemical degradation. Whereas the mechanical state in turn changes the fracture aperture. More recently, \citet{mollaali_variational_2025} generalized the approach by integrating finite-element reactive transport with a geochemical solver for capturing equilibrium and kinetic reactions. Porosity-dependent chemical damage is combined with phase-field degradation multiplicatively, while deformation and damage variables dynamically update the fluid storage capacity, fracture permeability, and solute diffusivity.
Along the same thread, \citet{borja_constitutive_2023} proposed a model based on the gradient-enhanced formulation, where chemical damage is related to the cumulative dissolved mineral mass, which assists the mechanical damage in terms of degrading the elastic modulus of dissolvable geomaterials. 
On the other hand, when precipitation takes place as the primary chemical reaction, the coupling effect could be modeled by introducing a precipitation-related eigenstrain that supplies the crack-driving energy, while the damage field feeds back by enhancing solute diffusivity~\citep{korec_phase-field_2023,xu_time-dependent_2025}.

Despite the success of the above-mentioned coupling schemes, they predominantly rely on the superposition of two distinct damage variables. As such, the chemical degradation is decoupled from mechanical damage and is treated as an independent factor acting on matrix stiffness or strain. Consequently, how the dissolution process alters the fracture tip topology and the physical dimension of the fracture process zone (FPZ) around the tip under stressed conditions is essentially not captured. Here, we propose a novel mechanism-based chemo-mechanical phase-field fracture framework that unifies mechanical and chemical degradation into a single phase-field variable. This integrated phase-field variable is achieved by considering the intrinsic evolution of the FPZ around a propagating crack-tip in a dissolvable rock under chemically reactive environment. Specifically, the model accounts for the reaction-diffusion of protons and the concurrent degradation of solid matrix due to mass removal of minerals, which alters the phase-field length scale, leading to an effective enlargement of the FPZ as the chemical dissolution progresses. A key advantage of this explicit coupling scheme lies in that the evolution of the FPZ is physically measurable, allowing the model to be calibrated with carefully designed experiments.

It is worth noting that treating the length scale as a characteristic dimension of FPZ is well supported by existing literature in gradient damage and variational phase-field models. For example, \citet{pham_gradient_2011} identified the internal length as an additional material parameter, and \citet{tanne_crack_2018} linked the length to the tensile strength of the material, demonstrating quantitative prediction of crack nucleation with various geometries. Recently, \citet{borja_da_rocha_damage_2026} adopted a finite internal length as a material property that sets the size of the damage process zone and demonstrated its role in governing the morphology of damage localization in soft solids. In the current work, our model takes one step forward in terms of transferring the static material length to a dynamic counterpart, which allows the physical size of the FPZ (representing the extent of micro-cracking) to evolve in response to the progressive chemical degradation of solid matrix under mechanical loading conditions.

To be specific, the proposed phase-field formulation intrinsically preserves the two-way coupling mechanism between chemical mass removal and matrix deformation developed for dissolvable rocks in~\citet{hu_environmentally_2013}, which is explained as follows. When irreversible deformation occurs, new micro-crack walls are generated, increasing the total Specific Surface Area (SSA) of the solid-fluid interface per unit volume. The chemical mass removal rate is thereby accelerated assuming it is linearly proportional to the SSA, which further weakens the solid matrix and enhances irreversible deformation. Therefore, a mutually amplified feedback between mechanical and chemical processes is established, which is most pronounced within the process zone around the crack-tip~\citep{hu_modeling_2019,tang_acid-assisted_2024}, where stress concentrates. Translated into the proposed phase-field model, the chemically induced evolution of the length scale alters the damage distribution around the crack-tip, which enhances ion diffusivity and reaction kinetics, gradually redefining the reaction-damage-diffusion affected zone and resulting in accelerated material degradation. 

The remainder of this paper is structured as follows. Section~\ref{sec:theory} establishes the theoretical framework, introducing the mechanisms of mineral dissolution and the reaction-diffusion formulation. Subsequently, the phase-field fracture formulation is derived via the crack surface approximation and the construction of the system’s Helmholtz potential energy. We highlight the physical interpretation of the phase-field length scale in the context of coupled chemo-mechanics, linking it to the extent of mineral mass removal to define the dissolution-damage coupling in degrading rocks. Section~\ref{sec:implement} details the numerical experiments, including the model validation, the setup of the initial-boundary value problem, and the convergence study. Section~\ref{sec:result} presents the numerical findings of our chemo-mechanical phase-field model, examining the effects of environmental pH and mechanical loading rates on crack propagation, placing a particular focus on their influence on the failure mode, the distribution of acidity, the evolution of mineral mass removal, and the stress response in the vicinity of the crack-tip. 

\section{Theory}
\label{sec:theory}
This section outlines the theoretical formulation developed in the present study. Specifically, the proposed model aims to address the following questions:
\begin{enumerate}
\item \textbf{FPZ evolution}: When a rock is exposed to a chemically reactive environment, how does mineral dissolution affect the internal length scale and reshape the FPZ around the tip of a propagating crack?

\item \textbf{Failure mode}: How does the interplay between chemical intrusion and mechanical loading determine the ultimate failure mode of a stressed rock matrix, given that it is undergoing a transition from brittle to more ductile response?

\item \textbf{Crack-tip stress}: How does the cross-scale chemo-mechanical feedback control the level of stress concentration ahead of the crack tip and its evolution under varying mechanical and chemical loading conditions?
\end{enumerate}

Motivated by these questions, Section~\ref{sec:mineral} establishes the mechanisms of mineral dissolution in a reactive environment at two distinct scales, incorporating one-way and two-way coupling, respectively. Subsequently, we introduce the reaction-diffusion equations in Section~\ref{sec:rd}, which specifically account for the feedback arising from the generation of fracture. For this purpose, a diffusive damage variable is employed through the phase-field approximation (Section~\ref{sec:phase_field}). 
A key novelty of this work lies in extending the physical meaning of the length scale of fracture, enabling the damage profile to evolve dynamically in response to the reactive environment.

\subsection{Mechanism of mineral dissolution}
\label{sec:mineral}
\begin{figure}[t]
\centering
\includegraphics[width=0.8\textwidth]{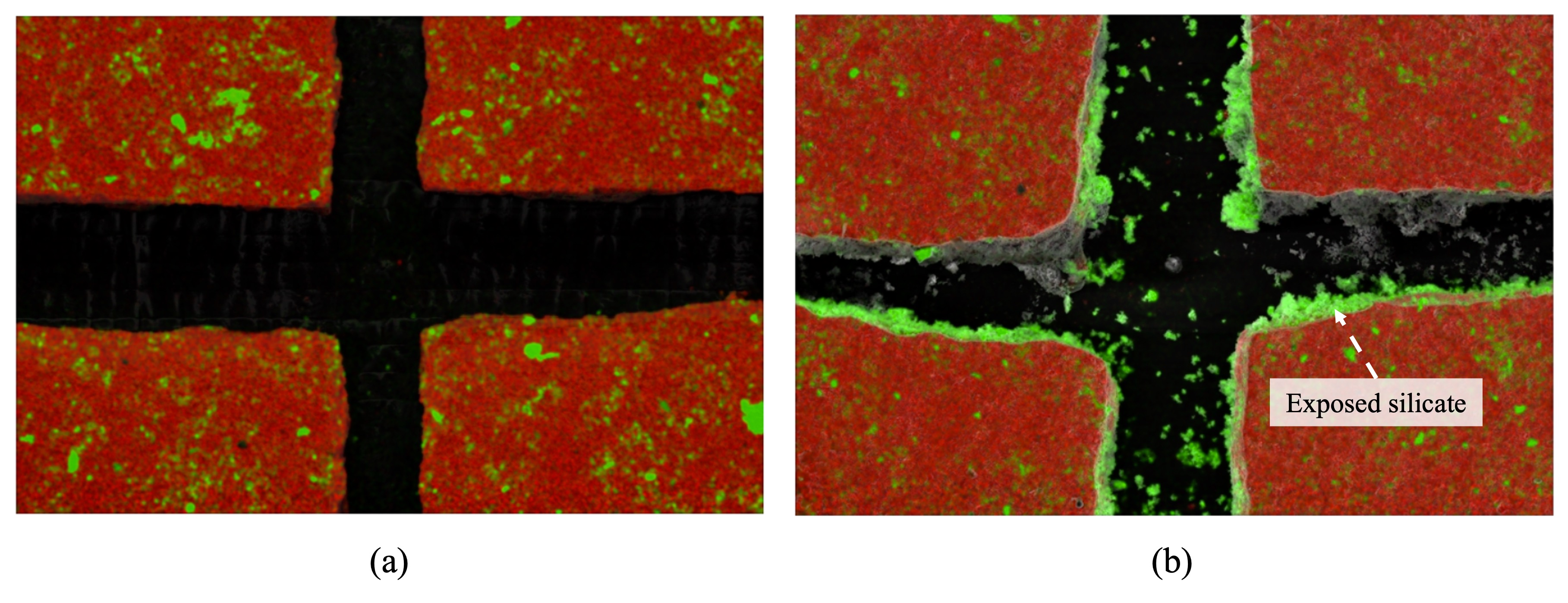}
\caption{Experimental evidence of mineral dissolution from microfluidic tests. SEM-based energy-dispersive spectroscopy images of carbonate-rich Marcellus shale samples: (a) pre-reaction condition, and (b) post-reaction condition after acidic fluid injection. Color codes: red for calcium and green for silicon. Figures adapted from \citet{ling_probing_2022}.}
\label{fig:exp_dissolution}
\end{figure}

Here, calcite dissolution is taken as a representative of the most common chemical processes in carbonate reservoirs. In weak acidic environments, the predominant chemical reaction involved in solid mass removal is described as:
\begin{equation}
\ce{CaCO_3}(\mathrm{s})+\ce{H^+}(\mathrm{aq}) \ce{<=>} \ce{Ca^{2+}}(\mathrm{aq})+\ce{HCO^{-}_{3}}(\mathrm{aq}) \,.
\label{eq:reaction}
\end{equation}
The above reaction represents the primary step in carbonate calcium-water interactions. While the aqueous bicarbonate product is unstable and tends to further react to form $\ce{CO_2}(\mathrm{g})$, we consider the reaction described in Eq.~\ref{eq:reaction} as the dominant process governing solid mass removal. As shown in Figure~\ref{fig:exp_dissolution}, scanning electron microscopy (SEM) images of microfluidic experiments have visualized the dissolution dynamics of carbonate rock upon injection of acid~\citep{ling_probing_2022}. It is clearly observed that the fracture channel morphology changed considerably after chemical mass removal of dissolvable minerals from the solid skeleton, leading to the exposure of nonreactive minerals.

Following \citet{hu_modeling_2019} and \citet{tang_reactive_2023}, the rate of calcite mass removal at the local scale (grain-scale) is defined as:
\begin{equation}
\dot{\xi}^{loc}=\beta_{\ce{H}^{+}} (C_{\ce{H}^{+}})^{k^{\prime}} \,,
\label{eq:mineral_dissolution_rate_loc}
\end{equation}
where $\beta_{\ce{H}^{+}}$ denotes the coefficient representing the effect of acidity on mass removal rate and $C_{\ce{H}^{+}}$ is the proton concentration. 
The exponential index $k^{\prime}$ is a constant parameter. Here, we adopt a linear relationship by setting the value to 1.0. The local chemical mass removal is then given by the integration of Eq.~\ref{eq:mineral_dissolution_rate_loc} over time:
\begin{equation}
\xi ^{loc}=\int \beta_{H^{+}} (C_{H^{+}})^{k^{\prime}} \mathrm{d}t \,.
\label{eq:mineral_dissolution_loc}
\end{equation}

At the Representative Elementary Volume (REV) scale, we consider that the rate of acidity-sensitive calcite dissolution also depends on the newly generated surface area due to microcracking~\citep{hu_environmentally_2013}. Leveraging the phase-field method, we establish the damage-enhanced mass removal rate dependent on a damage variable $d$, which will be detailed in Section~\ref{sec:phase_field}. Thus, the calcite dissolution rate at the REV scale reads:
\begin{equation}
\dot{\xi} ^{REV}= (1+\eta d) \beta_{\ce{H}^{+}}(C_{\ce{H}^{+}})^{k^{\prime}} \,,
\label{eq:mineral_dissolution_rate_REV}
\end{equation}
where $\eta$ is the coefficient of fracture enhancement on the process of chemical dissolution. By setting $\eta=0$, the mass removal rate at the REV scale is recovered to its local equivalent.

\subsection{Reaction-diffusion formulation}
\label{sec:rd}

\begin{figure}[t]
\centering
\includegraphics[width=0.5\textwidth]{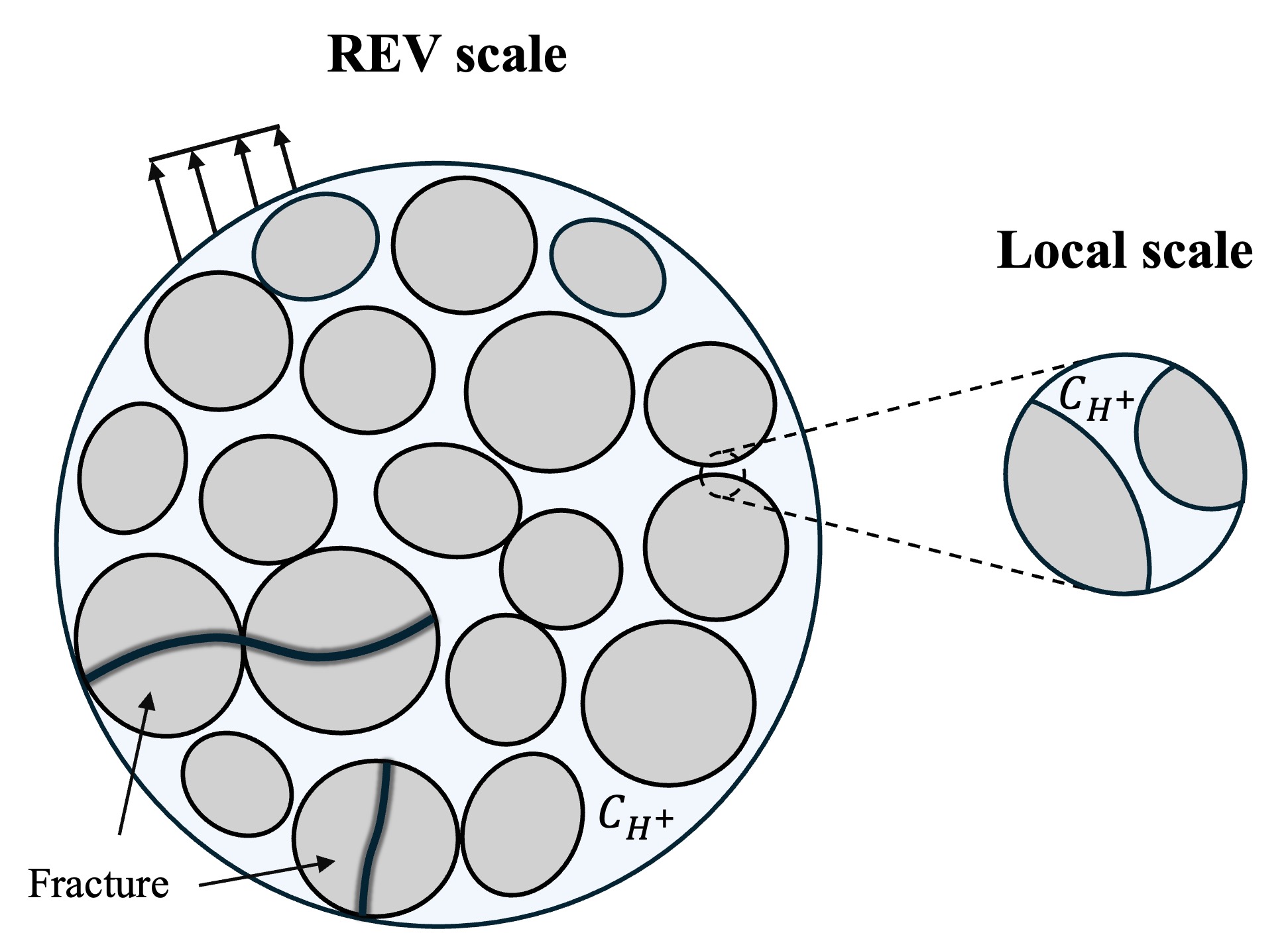}
\caption{Schematics representation of the REV scale and the local scale. The figure is an idealization of the microstructure of a carbonate rock. At the REV scale, the dissolution rate and solute diffusivity are coupled to the newly generated fracture pathways. While damage is not considered in the chemical model at the local scale.}
\label{fig:scale_schematics}
\end{figure}

The evolution of acidity described in Eq.~\ref{eq:reaction} can be formulated as a reaction-diffusion equation. As presented schematically in Figure~\ref{fig:scale_schematics}, our model distinguishes two characteristic scales. The first scale, referred to as the local scale, addresses sub-grain scale mineral reaction and its influence on the surrounding matrix. At this level, material damage is not considered and the rate of mass removal from the calcite dissolution process depends only on the delivery of proton (see Eq.~\ref{eq:mineral_dissolution_rate_loc}). The second scale is the REV scale, which captures the macroscopic behavior of geomaterial, incorporating the effect of damage and the formation of fracture pathways.

The reaction-diffusion equations are formulated at the REV scale. First, we couple the source term of the equation to the damage-enhanced mass removal rate $\dot{\xi}^{REV}$ (see Eq.~\ref{eq:mineral_dissolution_rate_REV}). Then, the diffusivity at the REV scale is also related to the damage variable to account for the enhanced diffusion process in the fully connected paths created by fracture. Thus, at the REV scale, the reaction-diffusion equation of solute $\ce{Ca}^{2+}$ is defined as:
\begin{equation}
\frac{\partial x_{\ce{Ca}^{2+}}}{\partial t} = D^{*}_{\ce{Ca}^{2+}} \nabla^2 x_{\ce{Ca}^{2+}} + \dot{\xi}^{REV} \,,
\end{equation}
where $x_{\ce{Ca}^{2+}}$ is the molar fraction of calcium ion and $D^{*}_{\ce{Ca}^{2+}}$ is the effective diffusivity of calcium ion.

In the same vein, we can formulate the reaction-diffusion processes of proton delivery as: 
\begin{equation}
\frac{\partial C_{\ce{H}^{+}}}{\partial t} = D^{*}_{\ce{H}^{+}} \nabla^2 C_{\ce{H}^{+}}-\gamma_{C H} \dot{\xi}^{REV} \,,
\label{eq:REV_reaction_diffusion}
\end{equation}
where $D^{*}_{\ce{H}^{+}}$ represents the effective diffusivity of proton. $\gamma_{CH}$ denotes a proportionality constant of the concentration change rate of proton due to consumption over the mass transfer rate of calcium production. 

The effective diffusivity of proton is formulated to incorporate the enhancement of diffusion associated with fracture development in the intact material. According to \citet{wu_phase-field_2016}, the effective diffusivity is defined as:
\begin{equation}
D^{*}_{\ce{H}^{+}} (d) = (1-d^m)D_0 + d^m D_c \,,
\end{equation}
where $D_0$ is the diffusivity of the undamaged material and $D_c$ denotes the intrinsic diffusivity of the species within a fully open crack. The parameter $m$ governs the rate of diffusivity evolution with respect to $d$. In this study, we select $m=5$ to model a diffusivity profile characterized by a non-linear transition: diffusion enhancement is negligible at low damage levels but increases rapidly as localized fracture forms. So far, in the reaction-diffusion equation of proton, the diffusion terms account for enhanced solute transport through fracture-induced pathways, while the source terms quantify the rate of mass flux resulting from damage-accelerated dissolution processes.

\subsection{Phase-field formulation}
\label{sec:phase_field}
In this section, we present a chemo-mechanical phase-field fracture modeling approach to address multiphysics challenges in crack propagation within acidized environments. Firstly, we introduce the phase-field variable for damage representation and the classical approximation of crack topology. The characteristic length scale, which governs the size of the FPZ, is discussed. Next, we formulate the potential energy of the system, comprising three components: the elastic strain energy stored in the solid, the energy dissipation from cracking, and the external energy sources. Subsequently, we establish the chemo-mechanical coupling scheme through a phase-field length scale that evolves dynamically with mineral mass removal. Using a variational approach, we derive the governing equations for momentum balance and crack evolution. Based on these formulations, we complete the multiphysics feedback loop between chemical dissolution and fracture propagation.

\subsubsection{Approximation of crack surface}
We first introduce the phase-field fracture theory, in which the discontinuous strain field is regularized by introducing a phase-field variable $d \in [0, 1]$. The phase-field variable is interpreted as material damage. $d = 1$ represents a completely damaged region, whereas $d = 0$ denotes an undamaged zone. The transition region $d\in(0,1)$ is regularized by an exponential function $d(x)=e^{-|x| / l_0}$ for brittle materials, with a characteristic length scale $l_0$. This exponential function arises as the solution to the following equation:
\begin{equation}
d - l^2_0 \nabla^2 d = 0 \,,
\end{equation}
which is the Euler–Lagrange equation of the variational principle
\begin{equation}
d=\operatorname{arg}\left\{\inf _{d \in W} \mathcal{L}(d)\right\} \,,
\end{equation}
where \(W=\{d \mid d(0)=1, d( \pm \infty)=0\}\) defines the space of admissible functions that satisfy boundary conditions for $d$. The energy functional is defined as:
\begin{equation}
\mathcal{L}(d)=\int F(x, d, \nabla d) \mathrm{d} V=\int\left(\frac{1}{2} d^2+\frac{l_0^2}{2} \nabla d \cdot \nabla d\right) \mathrm{d} V \,,
\label{eq:functional}
\end{equation}
where $F(x, d, \nabla d)$ represents the free energy density. The characteristic length scale $l_0$ plays a crucial role in determining the spatial extent of the transition region between different phases, such as the crack surface or the interface between two different materials~\citep{ambrosio_approximation_1990,amor_regularized_2009,kristensen_assessment_2021}. An increased value of $l_0$ results in a more diffusive interface, thus lowering the gradient energy term. In contrast, a smaller $l_0$ indicates a sharper transition with localized damage, reducing the bulk energy term.

Substituting $d(x)$ into the Eq.~\ref{eq:functional}, we rewrite the energy functional by the regularized crack surface $\Gamma_d$:
\begin{equation}
\mathcal{L}(d)=\int e^{-2|x| / l_0} \mathrm{d} V=\int_{-\infty}^{+\infty} e^{-2|x| / l_0} \Gamma_d \mathrm{d} x=l_0 \Gamma_d \,.
\end{equation}

Rearranging the above energy functional, we can obtain the regularized crack surface $\Gamma_d$:
\begin{equation}
\Gamma_d=\frac{\mathcal{L}(d)}{l_0} =\int\left(\frac{1}{2 l_0} d^2+\frac{l_0}{2} \nabla d \cdot \nabla d\right) \mathrm{d} V=\int \gamma(d, \nabla d) \mathrm{d} V \,,
\label{eq:gamma}
\end{equation}
where $\gamma$ represents the crack surface density function. This formulation employs the Allen–Cahn approximation for the crack surface, which describes a smooth transition from $d = 1$ along the crack to $d = 0$ away from the crack, and $l_0$ controls the width of the phase-field approximation zone.

\subsubsection{Helmholtz potential energy of the system}

\begin{figure}[t]
\centering
\includegraphics[width=0.9\textwidth]{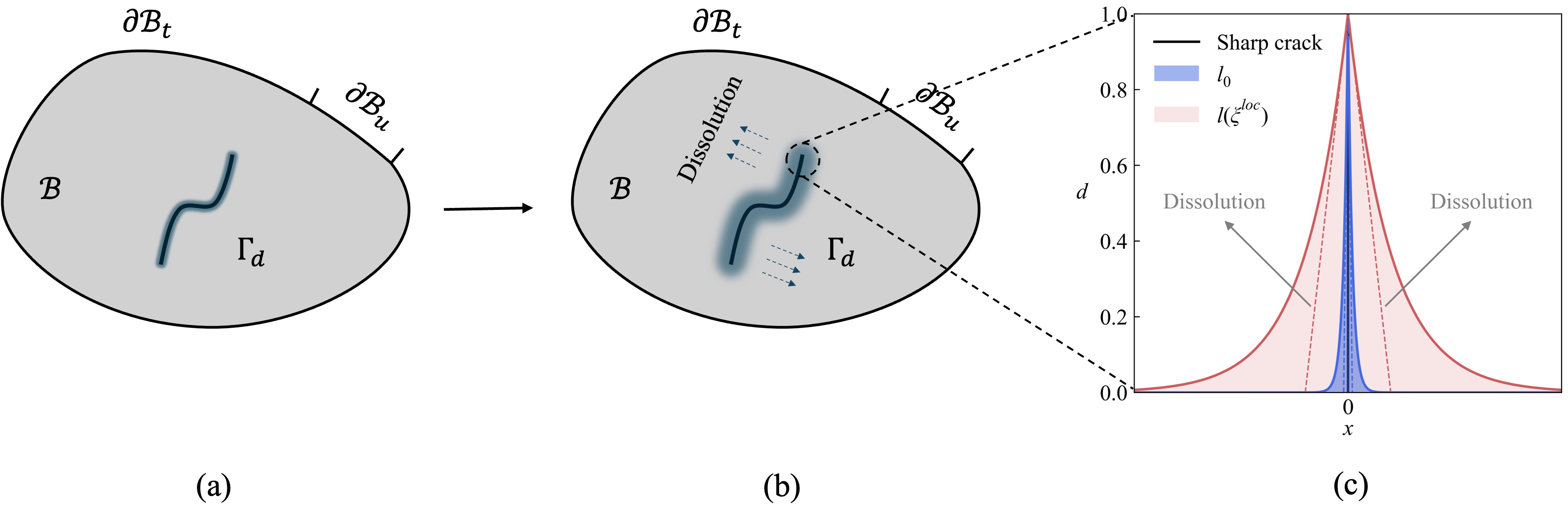}
\caption{Schematics of a solid body (a-b) with Dirichlet boundary $\partial \mathcal{B}_u$, Neumann boundary $\partial \mathcal{B}_t$, and an approximated crack surface area $\Gamma_d$: (a) crack surface before chemical dissolution, and (b) crack surface after chemical dissolution. Plot (c) shows the diffusive crack interface before and after dissolution.}
\label{fig:pf_schematics}
\end{figure}

Subsequently, consider a solid body $\mathcal{B}\subset \mathbb{R}^{dim}$ ($dim=2$ in this study) with boundary $\partial \mathcal{B}$, as illustrated in Figure~\ref{fig:pf_schematics}a. We decompose the whole boundary into the Dirichlet boundary $\partial \mathcal{B}_u$ and Neumann boundary $\partial \mathcal{B}_t$. The state of the system is described by the displacement field $\boldsymbol{u}$ and the phase-field $d$.

The total Helmholtz potential energy density of the system, denoted by $\psi_{\text {total}}$, is the sum of three distinct components in the presence of a crack:
\begin{equation}
\psi_{\text {total}} =  \psi_{\text {crack}} + \psi_{\text {internal}} + \psi_{\text {external}} \,,
\end{equation}
where $\psi_{\text {crack}}$ is the dissipation energy density of crack formation, $\psi_{\text {internal}}$ is the internal strain energy density stored from elastic deformation, and $\psi_{\text {external}}$ is the external energy density from traction and body force.

The phase-field method regularizes the strong discontinuities of the crack surfaces by introducing the aforementioned crack surface density function (Eq.~\ref{eq:gamma}). The dissipation energy density of crack formation can be expressed as follows:
\begin{equation}
\psi_{\text {crack}} = G_c \gamma(d, \nabla d) = G_c \left(\frac{1}{2 l_0}d^2+\frac{l_0}{2} \nabla d \cdot \nabla d\right) \,,
\end{equation}
where ${G}_c$ is the critical energy release rate. 

Assume linear isotropic elasticity, the elastic energy density in the intact solid domain is given by:
\begin{equation}
\psi_{\text {elastic}} (\boldsymbol{\varepsilon})=\frac{1}{2} \lambda \operatorname{tr}(\boldsymbol{\varepsilon})^2+\mu \operatorname{tr}(\boldsymbol{\varepsilon}^2) \,,
\end{equation}
where $\lambda$ and $\mu$ are the Lamé moduli. In a small deformation setting, the infinitesimal strain tensor is:
\begin{equation}
\boldsymbol{\varepsilon} = \frac{1}{2}\left(\nabla \boldsymbol{u}+\nabla \boldsymbol{u}^T\right) \,.
\end{equation}

To couple phase-field fracture with the degradation of mechanical behaviors, the concept of damage mechanics is adopted by incorporating a degradation function $g(d)$ into the internal energy function. $g(d)$ varies from 1 to 0 as the material transitions from completely intact $(d=0)$ to fully fractured $(d=1)$, capturing the progressive stiffness reduction associated with damage evolution. In this work, we employ the quadratic polynomial as the degradation function, which is expressed as: 
\begin{equation}
g(d)=(1-d)^2(1-k)+k \,,
\label{eq:degradation}
\end{equation}
where a small value of parameter $k$ is chosen to prevent the elastic energy density from vanishing and causing numerical singularity when $d$ approaches 1 \citep{bilgen_crack-driving_2019,ip_modeling_2023}.

Next, the elastic stored energy density is split into an active component coupled with phase-field fracture and an inactive part independent of fracture. Here, we adopt the spectral strain decomposition to decompose the elastic stored energy density into tension and compression parts, ensuring that degradation occurs only in tensile regions~\citep{miehe_thermodynamically_2010-1}. Consequently, the internal energy density can be formulated as follows:
\begin{equation}
\psi_{\text {internal}} = g(d) \psi_{\text{elastic}}^{+}(\boldsymbol{\varepsilon}) + \psi_{\text{elastic}}^{-}(\boldsymbol{\varepsilon}) \,,
\end{equation}
where $\psi_{\text {elastic}}^{+}(\boldsymbol{\varepsilon})$ and $\psi_{\text {elastic}}^{-}(\boldsymbol{\varepsilon})$ represent the tensile and compressive contributions to the elastic energy density, respectively. The stress tensor can be derived from the internal energy density as follows:
\begin{equation}
\boldsymbol{\sigma} = g(d) \frac{\partial \psi_{\text{elastic}}^{+} (\boldsymbol{\varepsilon})}{\partial \boldsymbol{\varepsilon}} + \frac{\partial \psi_{\text{elastic}}^{-} (\boldsymbol{\varepsilon})}{\partial \boldsymbol{\varepsilon}} \,.
\end{equation}

Next, the external energy density is given by:
\begin{equation}
\psi_{\text{external}}=- \boldsymbol{\tau} \cdot \boldsymbol{u} - \boldsymbol{b} \cdot \boldsymbol{u} \,,
\end{equation}
where $\boldsymbol{\tau}$ is the prescribed traction and $\boldsymbol{b}$ is the body force. For the sake of simplicity, gravitational force is not considered in this work.

Therefore, the total Helmholtz potential energy density can be written as the sum of the above three components:
\begin{align}
\psi_{\text {total}} = & \underbrace{G_c\left(\frac{1}{2 l_0}d^2+\frac{l_0}{2} \nabla d \cdot \nabla d\right)}_{\text {fracture energy}} + \underbrace{g(d) \psi_{\text {elastic}}^{+}(\boldsymbol{\varepsilon}) + \psi_{\text {elastic}}^{-}(\boldsymbol{\varepsilon})}_{\text {mechanically degraded internal energy}}  \nonumber\\
&- \underbrace{ (\boldsymbol{\tau} \cdot \boldsymbol{u} + \boldsymbol{b} \cdot \boldsymbol{u})}_{\text {external energy }} \,.
\end{align}

\subsubsection{Chemo-mechanical coupling}
\label{sec:length_coupling}
As discussed above, the phase-field length scale determines the spatial extent of the transition zone between the fully damaged and the completely intact region. In a non-reactive setting where only mechanical loading is considered, this length scale is typically treated as a constant material property. When subject to the reactive environment, the transition zone, characterized by micro-crack formation, develops due to mass loss triggered by chemical intrusion (as visualized in Figure~\ref{fig:exp_dissolution}). To address this, we propose a novel chemo-mechanical coupling approach that dynamically links the characteristic length scale of phase-field fracture to the evolving chemical processes. In this formulation, the length scale $l$ is governed by the local mass removal $\xi^{loc}$ (defined in Eq.~\ref{eq:mineral_dissolution_loc}), representing a chemical degradation effect. For brevity, we refer to $\xi^{loc}$ simply as mass removal hereafter. As dissolution progresses, the enhanced chemical damage around the crack tip, manifested through micro-crack development, induces a corresponding increase in the length scale $l$, as illustrated in Figures~\ref{fig:pf_schematics}a-b. This coupling mechanism allows the FPZ to evolve in response to the reaction-diffusion processes and the corresponding mass removal, effectively capturing the dynamic interplay between chemical dissolution and fracture propagation. The proposed model provides a mechanism-based method for modeling reactive fracture processes where chemical and mechanical damage mechanisms are intrinsically interdependent. In this work, the relation between the phase-field length scale and the dimensionless mass removal is defined as:
\begin{equation}
l(\xi^{loc}) = (1+\alpha \xi^{loc})l_0 \,,
\label{eq:length}
\end{equation}
where $\alpha$ is a constant coefficient determining the extent of mass removal enhancement on the phase-field length scale. The linear form is adopted as the leading-order term of the growth of the FPZ based on laboratory data. Specifically, the coefficient $\alpha$ in this work is calibrated using limestone samples under dissolution in a three-point bending test detailed in Section~\ref{sec:validation}. The predicted width of the FPZ under various acidic conditions is subsequently validated with experimental data detailed in Section~\ref{sec:width_fpz}.
It is emphasized that the proposed coupling with the chemical mass removal progressively taking place from the solid matrix preserves the $\Gamma$-convergence of the phase-field functional~\citep{bourdin_variational_2008}. The coupled length scale remains proportional to $l_0$, thereby recovering the Griffith sharp-crack limit as $l_0 \to 0$.

To further illustrate the physical interpretation of the proposed coupling scheme, Figure~\ref{fig:pf_schematics}c presents the one-dimensional profiles of phase-field $d$ perpendicular to the fracture surface before and after mineral dissolution. The dissolution-enhanced transition state of phase-field is given by:
\begin{equation}
d(x)=e^{-|x| / l(\xi^{loc})} \,.
\end{equation}
The profile widens under the influence of mineral dissolution compared to the reference state. The degradation function in Eq.~\ref{eq:degradation} is now capable of capturing the chemo-mechanical damage induced by both mechanical loading and mineral dissolution. Therefore, the total Helmholtz potential energy density can be rewritten as:
\begin{align}
\psi_{\text {total}} = & \underbrace{G_c \left(\frac{1}{2 l(\xi^{loc})}d^2+\frac{l(\xi^{loc})}{2} \nabla d \cdot \nabla d\right)}_{\text {chemically altered fracture energy}} + \underbrace{ g(d) \psi_{\text {elastic}}^{+}(\boldsymbol{\varepsilon}) + \psi_{\text {elastic}}^{-}(\boldsymbol{\varepsilon})}_{\text {chemo-mechanically degraded internal energy}}  \nonumber\\
&- \underbrace{( \boldsymbol{\tau} \cdot \boldsymbol{u} + \boldsymbol{b} \cdot \boldsymbol{u} )}_{\text {external energy}} \,.
\end{align}

Finally, the system can be solved by the minimization of the total potential energy. Following the approach of \citet{francfort_revisiting_1998,bourdin_variational_2008}, the variational formulation of the total potential energy is expressed as:
\begin{equation}
\nabla \cdot \frac{\partial \psi_{\text {total}}}{\partial \nabla \boldsymbol{u}} - \frac{\partial \psi_{\text {total}}}{\partial \boldsymbol{u}} = \mathbf{0} \,,
\label{eq:strong1}
\end{equation}
\begin{equation}
\nabla \cdot \frac{\partial \psi_{\text {total}}}{\partial \nabla d} - \frac{\partial \psi_{\text {total}}}{\partial d} = 0 \,.
\label{eq:strong2}
\end{equation}

By deriving the Fréchet derivative terms with respect to $\boldsymbol{u}$ and $d$, the minimization problem’s strong form can be obtained. Substituting the total potential energy in Eq.~\ref{eq:strong1}, we obtain the macroscopic momentum balance:
\begin{equation}
\nabla \cdot \boldsymbol{\sigma} + \boldsymbol{b} =\mathbf{0} \,.
\label{eq:momentum_balance}
\end{equation}

Substituting the total potential energy in Eq.~\ref{eq:strong2}, we have the fracture evolution equation:
\begin{equation}
\nabla \cdot \left(G_c l(\xi^{loc}) \nabla d\right) - \frac{G_c}{l(\xi^{loc})} d = g^\prime(d) \psi_{\text {elastic}}^{+}(\boldsymbol{\varepsilon}) \,.
\label{eq:fracture_evo}
\end{equation}

The boundary conditions are thereby given as:
\begin{equation}
\label{eq:boundary_condition}
\left\{\begin{array}{l}
\boldsymbol{u}=\boldsymbol{u}_0, \text { on } \partial \mathcal{B}_u \\
\boldsymbol{\sigma} \cdot \boldsymbol{n}=\boldsymbol{\tau}, \text { on } \partial \mathcal{B}_t \\
\nabla d \cdot \boldsymbol{n}=0, \text { on } \partial \mathcal{B}
\end{array}\right. \,.
\end{equation}

In Eq.~\ref{eq:fracture_evo}, crack propagation is driven by the elastic strain energy. However, crack healing may occur if the elastic strain energy decreases, thereby violating the irreversibility condition of the phase-field $d$. We introduce a historical variable $\mathcal{H}$ to enforce irreversibility as proposed by~\citet{miehe_thermodynamically_2010-1}. $\mathcal{H}$ is updated according to the following criterion: At each time increment, if $\mathcal{H}^{(t+\Delta t)} > \mathcal{H}^{(t)}$, we adopt the new value of $\mathcal{H}^{(t+\Delta t)}$; otherwise, the previous value $\mathcal{H}^{(t)}$ is retained. It corresponds to taking the maximum value of the elastic energy over time. Therefore, the new driving force of phase-field fracture takes the following form:
\begin{equation}
\label{eq:history-variable}
\mathcal{H}^{(t+\Delta t)} = \max_{\tau\in(0,t)} \psi_{\text {elastic}}^{+} (\boldsymbol{\varepsilon}) \,.
\end{equation}

\section{Numerical implementation and experiment}
\label{sec:implement}
This section outlines the numerical framework in this study. The governing equations are solved with a staggered computational scheme. Within each load step, the chemo-elasticity and fracture evolution subproblems are solved iteratively. The scheme is implemented using RACCOON, an open-source parallel finite element code for phase-field fracture built on the MOOSE framework~\citep{permann_moose_2020,Raccoon2020}.

In Section~\ref{sec:benchmark}, we first validate our phase-field implementation via a widely used benchmark case. Section~\ref{sec:validation} then validates the chemo-mechanical coupling scheme against three-point bending experiments under concurrent chemical and mechanical loading. Section~\ref{sec:prob_statement} defines the specific initial boundary value problem, a Mode I tensile crack subjected to a reactive environment, enabling parametric analysis of chemo-mechanical coupling as the fracture propagates. Lastly, we conduct a convergence study on the mesh and timestep sensitivity in Section~\ref{sec:convergence}.

\subsection{Validation of phase-field fracture implementation}
\label{sec:benchmark}

\begin{figure}[t]
    \centering
    \includegraphics[width=0.9\textwidth]{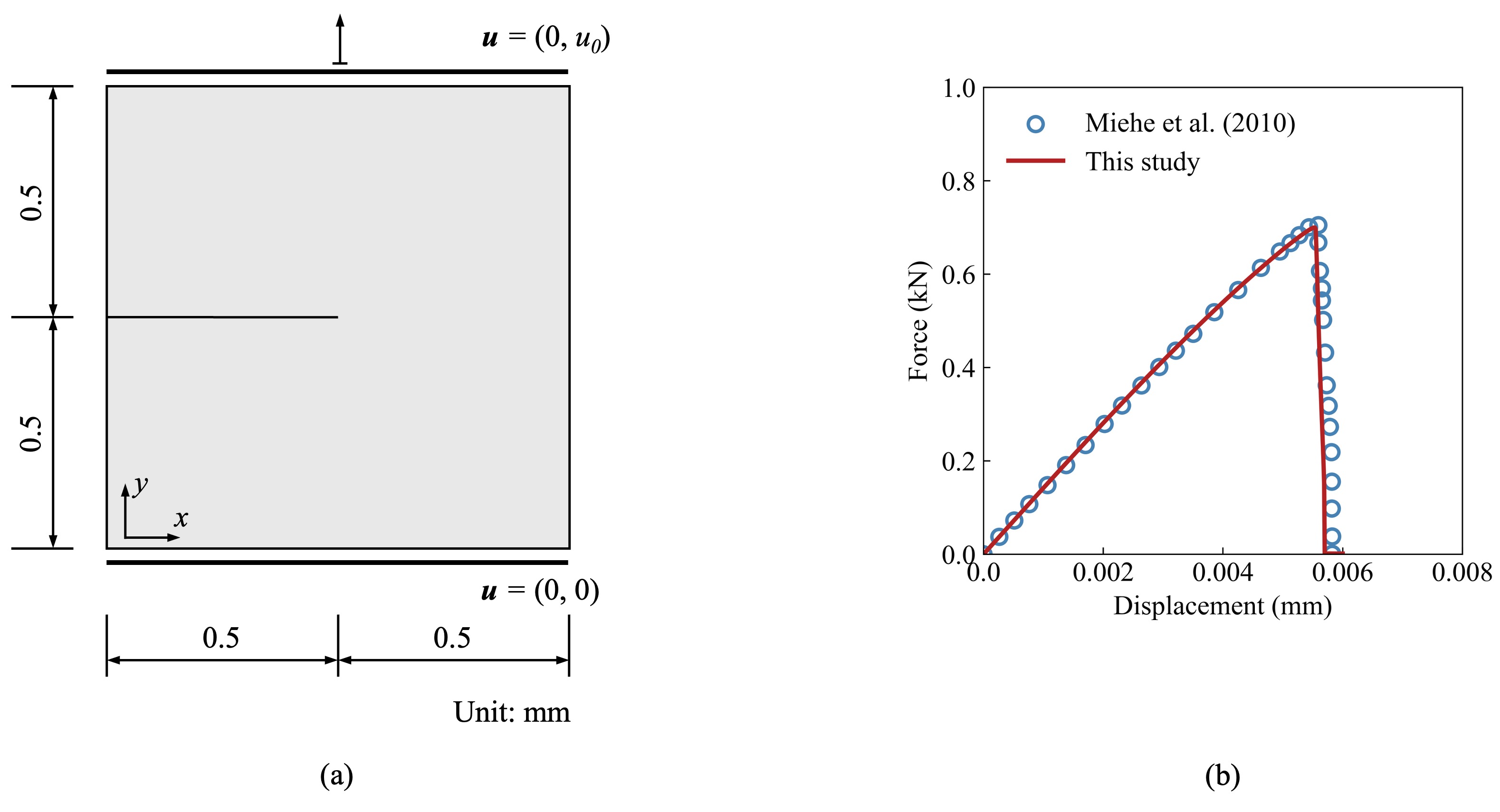}
    \caption{(a) Geometry and boundary conditions for the benchmark case. (b) Load--displacement curve of the single edge notched tension test, with comparison against the results from \cite{miehe_phase_2010}.}
    \label{fig:benchmark}
\end{figure}

Prior to investigating the concurrent chemo-mechanical effects, we validate the numerical implementation of the phase-field fracture method by simulating a single-edge notched tension test, a standard benchmark established by \cite{miehe_phase_2010}. The geometry and boundary conditions of the pre-notched square specimen are illustrated in Figure~\ref{fig:benchmark}a. The specimen is subjected to displacement-controlled vertical loading at the top edge, while the bottom edge remains fully constrained in both the horizontal and vertical directions. The loading is applied up to a total displacement of $6 \times 10^{-3}$ mm using fixed increments of $1 \times 10^{-5}$ mm. Consistent with the parameters used by \cite{miehe_phase_2010}, the material properties are defined as: Young’s modulus $E = 210$ GPa, Poisson’s ratio $\nu = 0.3$, critical energy release rate $G_c = 2.7$ N/mm, and phase-field regularization length $l = 0.015$ mm. Figure~\ref{fig:benchmark}b compares the load--displacement response obtained from our simulation with the benchmark results. Our result demonstrates an excellent agreement with the reference case, accurately capturing both the peak force and the subsequent rapid crack propagation.

\subsection{Validation of the chemo-mechanical coupling}
\label{sec:validation}

\begin{figure}[t]
    \centering
    \includegraphics[width=0.9\textwidth]{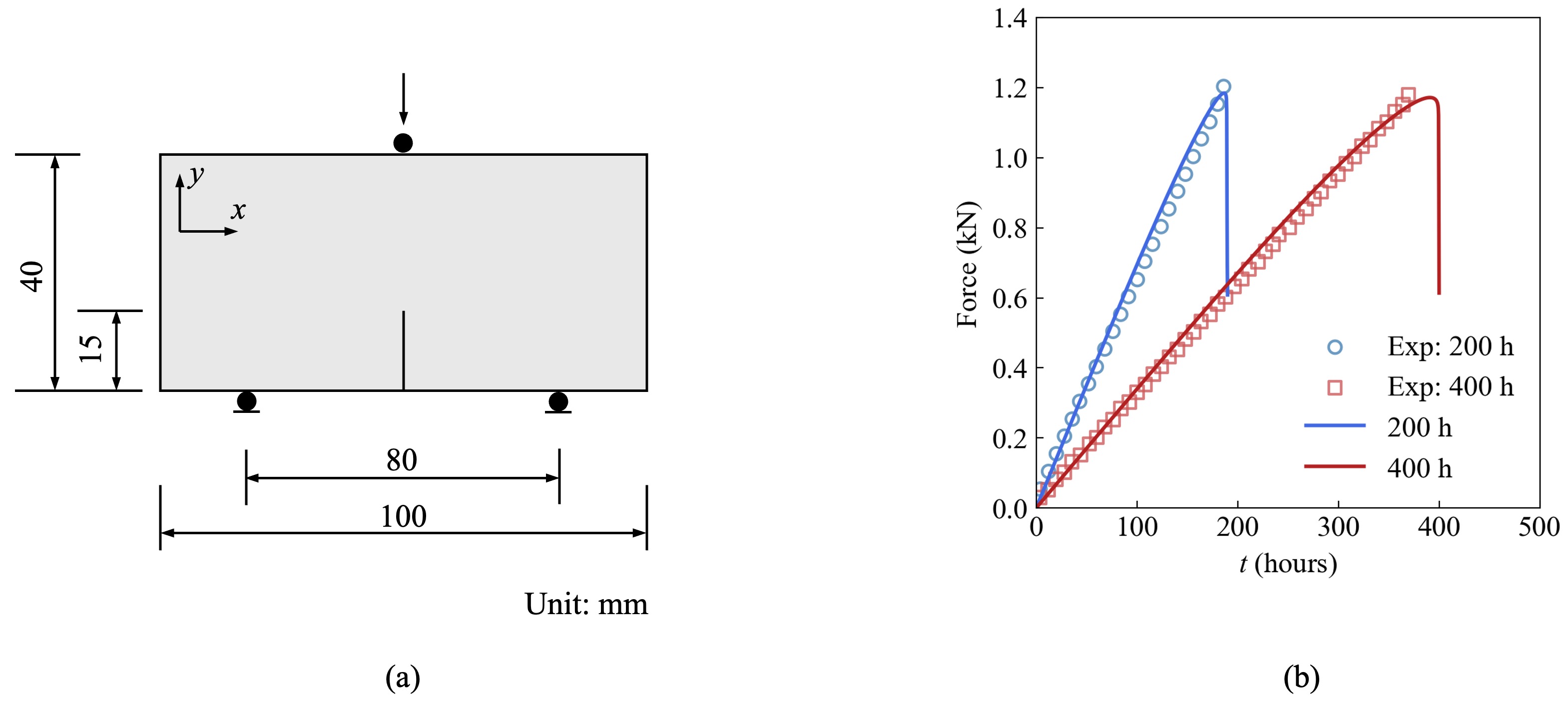}
    \caption{(a) Geometry and boundary conditions for the three-point bending tests on limestone immersed in an acidic solution. (b) Comparison between the predicted evolution of the reaction force and the experimental results under two chemo-mechanical loading conditions.}
    \label{fig:validation}
\end{figure}

To validate the proposed chemo-mechanical coupling scheme and to calibrate the coupling parameters in Eq.~\ref{eq:length}, we model a three-point bending test under concurrent chemical dissolution and mechanical loading and compare with the experiments conducted in this study. In the experiments, limestone specimens with calcite content above 93\% are used. The geometry and boundary conditions are illustrated in Figure~\ref{fig:validation}a. The beam measures $100 \times 40$ mm, with a support span of 80 mm and a 15 mm pre-notch crack at the middle of the sample. During loading, each specimen was immersed in a hydrochloric acid solution with a pH value of 3 in a container at room temperature. A slow loading rate was applied to ensure a prolonged exposure time for the relatively slow chemical processes. Accordingly, two loading durations were investigated, bringing the specimens to failure after approximately 200 and 400 h, respectively. Figure~\ref{fig:validation}b compares the predicted evolution of the reaction force with the experimental results at both loading rates. The calibrated model predicts the peak load and time to failure for the degraded limestone samples. Notably, the model captures the observed rate dependence, where a longer period of chemical exposure leads to a 1.8\% decrease in the peak load. The mechanical and chemical parameters employed are reported in Table~\ref{tab:parm}.

\subsection{Problem statement}
\label{sec:prob_statement}

\begin{figure}[t]
    \centering
    \includegraphics[width=0.8\textwidth]{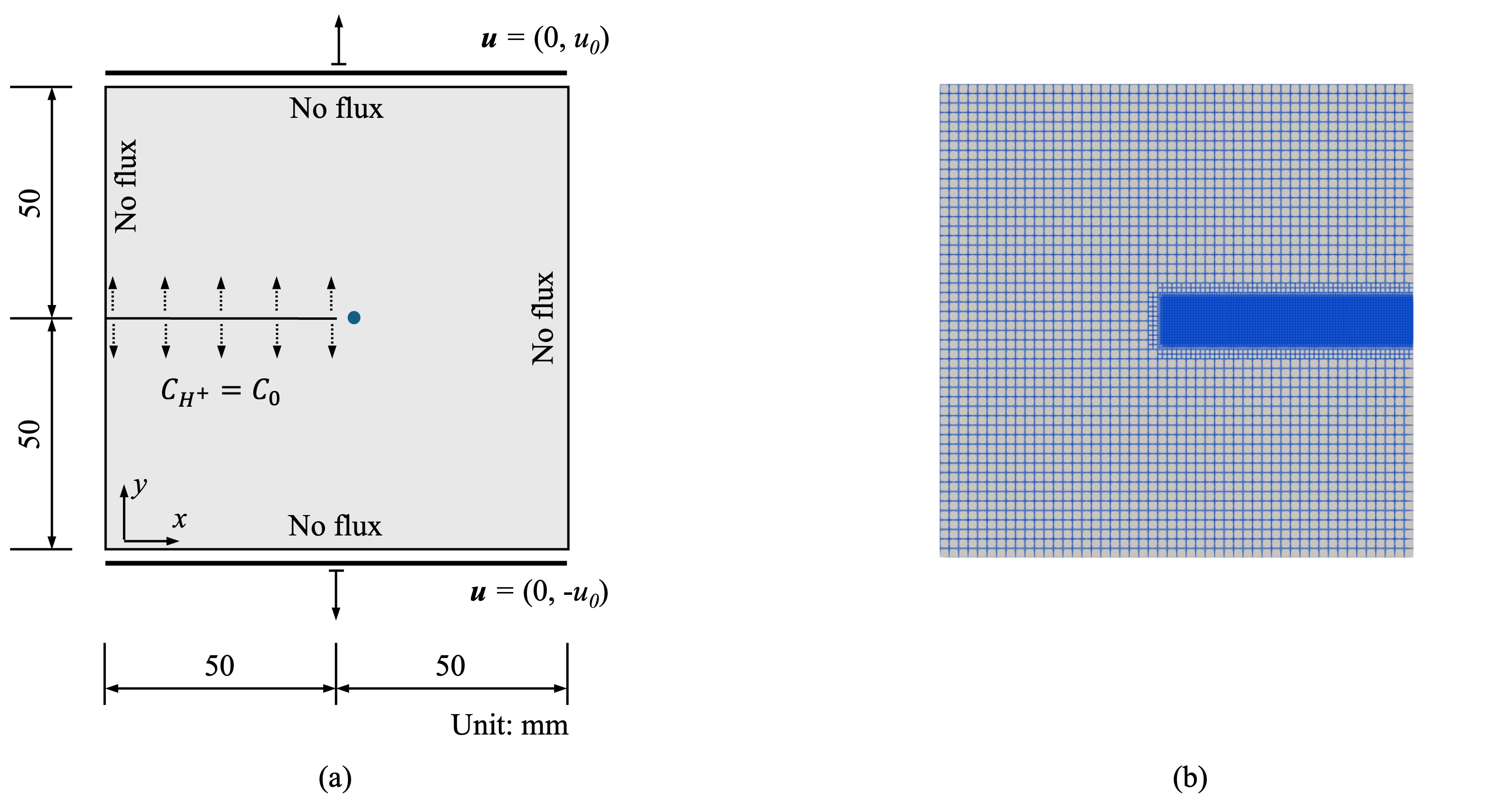}
    \caption{(a) Geometry and boundary conditions of the sample investigated in this study. The blue dot 2 mm ahead of the initial crack marks the position for damage evolution monitoring. (b) Local mesh refinement on the crack path.}
    \label{fig:geometry}
\end{figure}

Figure~\ref{fig:geometry} shows the setup of the initial--boundary value problem for investigating the chemo-mechanical coupling fracture behaviors in this study. The solid domain measures $100 \times 100$ mm and contains a 50 mm long horizontal crack located at the left-center of the specimen, which is initially generated through geometric discontinuity. The domain is discretized using a $50 \times 50$ base grid, with local mesh refinement applied around the crack tip and along the anticipated propagation path. Carbonate-rich limestone was selected as the target material due to its high reactivity in acidic environments. A source of constant proton concentration, $C_0$, is imposed on the surface of the pre-existing crack, while no-flux boundary conditions are applied to the outer boundary of the domain. Three acidity levels, corresponding to pH values of 3.0, 3.5, and 4.0, are investigated in Section~\ref{sec:pH_effect}. The remaining solid domain is initialized with a proton concentration corresponding to the neutral pH condition. The top and bottom edges are subjected to displacement-controlled Mode-I tensile loading. A baseline loading rate of $v = 2.5 \times 10^{-10}$ s$^{-1}$ is used. The influence of mechanical loading rate is subsequently discussed in Section~\ref{sec:loading_rate_effect}. More validation cases of the predicted width of the FPZ against experimental data are detailed in Section~\ref{sec:width_fpz} of the results.

We note that the present model targets the diffusion-dominated low-permeability material, in which species transport is governed by diffusion and advection is considered secondary. 

\begin{table}
\centering
\caption{Summary of mechanical and chemical parameters.}
\begin{tabular}{l c c c}
\hline
        Parameter & Symbol & Value & Unit  \\
\hline
        \multicolumn{4}{l}{\textbf{Mechanical}} \\
        Young's modulus  &$E$  & 25 & GPa \\
        Poisson's ratio  &$\nu$ & 0.25 & -  \\
        Critical energy release rate &$G_c$ & 0.03 & N/mm  \\
        Initial regularization length &$l_0$ & 1 & mm  \\
\hline        
        \multicolumn{4}{l}{\textbf{Chemical}} \\
        Diffusivity of undamaged material &$D_0$ & $1.25 \times 10^{-4}$ & mm$^2$/s  \\
        Diffusivity of crack opening &$D_c$ & $3.6 \times 10^{-3}$ & mm$^2$/s \\
        Constant denoting the effect of acidity on mass removal rate &$\beta_{\ce{H}^{+}}$ & $2.5 \times 10^{-3}$ & s$^{-1}$ \\
        Coefficient of fracture enhancement on dissolution &$\eta$ & $1 \times 10^{3}$ & - \\
        Extent of mass removal enhancement on fracture length scale &$\alpha$ & $2 \times 10^{3}$ & - \\
\hline
\end{tabular}
\label{tab:parm}
\end{table}

\subsection{Convergence study}
\label{sec:convergence}
\begin{figure}[t]
    \centering
    \includegraphics[width=0.8\textwidth]{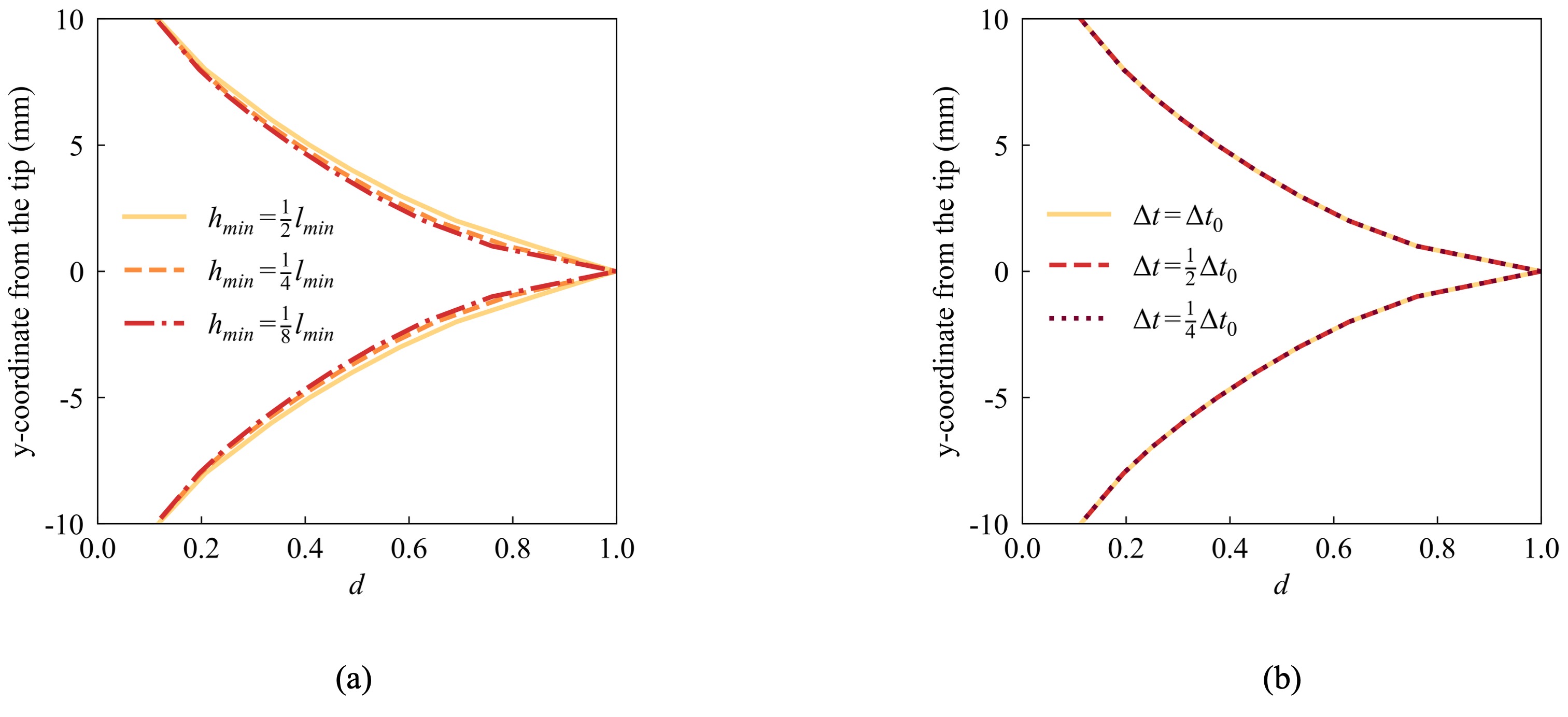}
    \caption{Convergence study: (a) mesh size, (b) timestep size.}
    \label{fig:convergence}
\end{figure}

To ensure numerical reliability of the proposed model, a convergence study on both the mesh size and the timestep size were first conducted. For this analysis, the reference case subject to the most reactive environment (pH = 3.0) under the baseline tensile loading rate is employed. Figure~\ref{fig:convergence} shows the phase-field damage distribution perpendicular to the crack path at the center of the specimen ($x = 50$ mm) at the crack onset, considering prescribed spatial and temporal discretizations.

The minimum phase-field length scale is fixed at $l_{min} = 1$ mm. Then, we progressively increased the mesh refinement around the crack tip, thereby decreasing the minimum element size, $h_{min}$. Figure~\ref{fig:convergence}a demonstrates that the damage profile converges as the mesh is refined. An element size of $h_{min} = \frac{1}{8} l_{min}$ is found sufficient to resolve the sharp damage gradient at the crack tip. Subsequently, we fixed the mesh refinement level at $h_{min} = \frac{1}{8} l_{min}$ to investigate the sensitivity of the results to the timestep size. Using $\Delta t_0 = 1000$ s as a base case, we reduced the timestep to $\Delta t = \frac{1}{2} \Delta t_0$ and $\Delta t = \frac{1}{4} \Delta t_0$, respectively. As shown in Figure~\ref{fig:convergence}b, the damage profile remains consistent across these variations. These results confirm that the timestep $\Delta t_0$ is sufficiently small to capture the damage evolution within the context of coupled chemo-mechanics under concern.

\section{Results}
\label{sec:result}
This section presents the numerical results obtained from the proposed framework. We begin with presenting the characteristics of reactive fracture propagation compared with purely mechanically driven fracture in Section~\ref{sec:chemo_mechanical}. Next, we investigate the two primary factors influencing fracture behaviors under the chemo-mechanical conditions, namely, the environmental acidity (i.e., pH) in Section~\ref{sec:pH_effect} and mechanical loading rate in Section~\ref{sec:loading_rate_effect}. Section~\ref{sec:stress_resp} presents the crack-tip stress responses under environmental constraints of various acidity levels and loading rates. Finally, Section~\ref{sec:width_fpz} compares the predicted width of FPZ in chemically reactive environments with experimental results.

\subsection{Chemo-mechanical fracture propagation}
\label{sec:chemo_mechanical}
\begin{figure}[t]
    \centering
    \includegraphics[width=\textwidth]{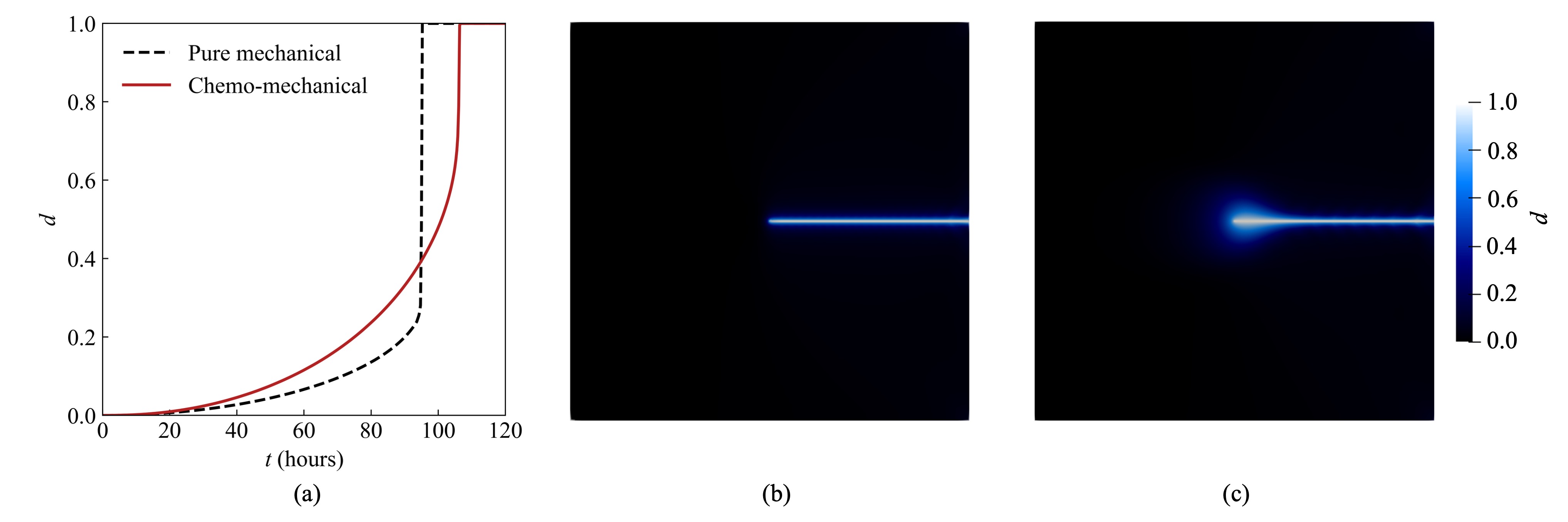}
    \caption{(a) Comparison of the evolution of phase-field damage at the front of the pre-existing crack tip in pure mechanical and chemo-mechanical cases. The tracking point is marked in blue in Figure~\ref{fig:geometry}. Distribution of phase-field damage at 120 h in samples under (b) pure mechanical loading and (c) chemo-mechanical loading.}
    \label{fig:comparison}
\end{figure}

We first compare the case of crack propagation under chemo-mechanical influence using our phase-field fracture model with a purely mechanically driven counterpart. Figure~\ref{fig:comparison}a illustrates the evolution of phase-field damage at the front of the pre-existing crack tip for these two scenarios. In the purely mechanical case, where only tensile loading is applied to the sample and no chemical effects are considered, the fracture length scale is constant and independent of the surrounding environment. At the crack tip, damage accumulates gradually with the applied mechanical load, followed by a sharp increase in the damage variable, $d$, towards a value of 1.0, indicating brittle fracture propagation. The increase in $d$ from a small value to a completely fractured state occurs within a short period of time. In contrast, the fracture evolution in the chemo-mechanical setting exhibits a more ductile behavior. Here, the pre-existing notch is subjected to external tensile loading while simultaneously being exposed to an acidic environment with a pH of 3.0. A more gradual increase in phase-field damage is observed, resulting in a significant delay in reaching the completely fractured state where d approaches 1.0. Although the initial deterioration at the crack tip is more severe in the chemo-mechanical case due to chemical mass removal, in the purely mechanical case the fracture extends in a more brittle manner, and its accumulated damage exceeds that of the chemo-mechanical case at approximately 95 h.

Figures~\ref{fig:comparison}b and \ref{fig:comparison}c present the phase-field damage profiles at 120 h for the purely mechanical and chemo-mechanical cases, respectively. In the purely mechanical case, as Figure~\ref{fig:comparison}b displays, a sharp Mode-I crack profile with a constant crack width nucleates at the pre-existing tip and propagates toward the right side of the sample. In contrast, as shown in Figure~\ref{fig:comparison}c, the crack propagation under chemo-mechanical influence reveals a diffusive and enlarged FPZ around the pre-existing crack tip, where stress concentrates and localized chemical degradation takes place concurrently. The non-uniform damage distribution is attributed to the varying degrees of chemical exposure and material deterioration.

\begin{figure}[t]
    \centering
    \includegraphics[width=\textwidth]{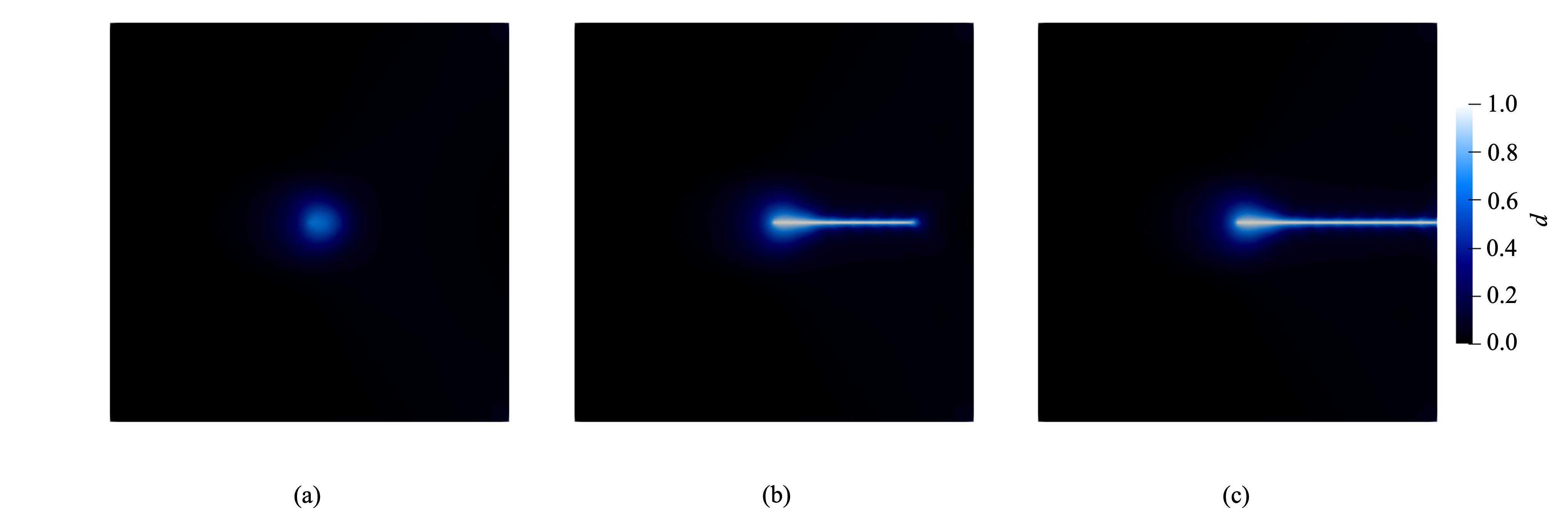}
    \caption{Evolution of chemo-mechanical cracking: (a) initial stage of crack propagation, (b) during crack propagation, and (c) crack propagation completed.}
    \label{fig:evolution}
\end{figure}

Time evolution of the phase-field fracture during propagation under coupled chemo-mechanical feedback using the proposed model is illustrated in Figure~\ref{fig:evolution}. Under the combined effect of tensile loading and environmental loading by the diffusion of acidic agents from the pre-existing crack, damage accumulates around the crack-tip. Upon the initial extension of the crack, a small rounded process zone, characterized by a diffusive damage profile, forms around the crack-tip. When the localized stress at the tip reaches the threshold, the crack starts to propagate along the direction perpendicular to the tensile loading. As the characteristic timescale for the crack propagation rate is much smaller than that of the diffusion of reactive agents, the newly generated crack surface remains relatively sharp, forming a region herein referred to as a ``chemical-lag zone". Therefore, the crack profile can be divided into two distinct regions: in the initial chemical-filled zone, the crack width depends on the local extent of chemical mass removal, whereas in the subsequent chemical-lag zone, the material remains unaffected by the acidic environment and thus exhibits a relatively constant crack width.

\subsection{The effect of environmental pH on crack propagation}
\label{sec:pH_effect}

\subsubsection{Brittle-to-ductile transition with decreasing environmental pH}
\begin{figure}[t]
    \centering
    \includegraphics[width=0.8\textwidth]{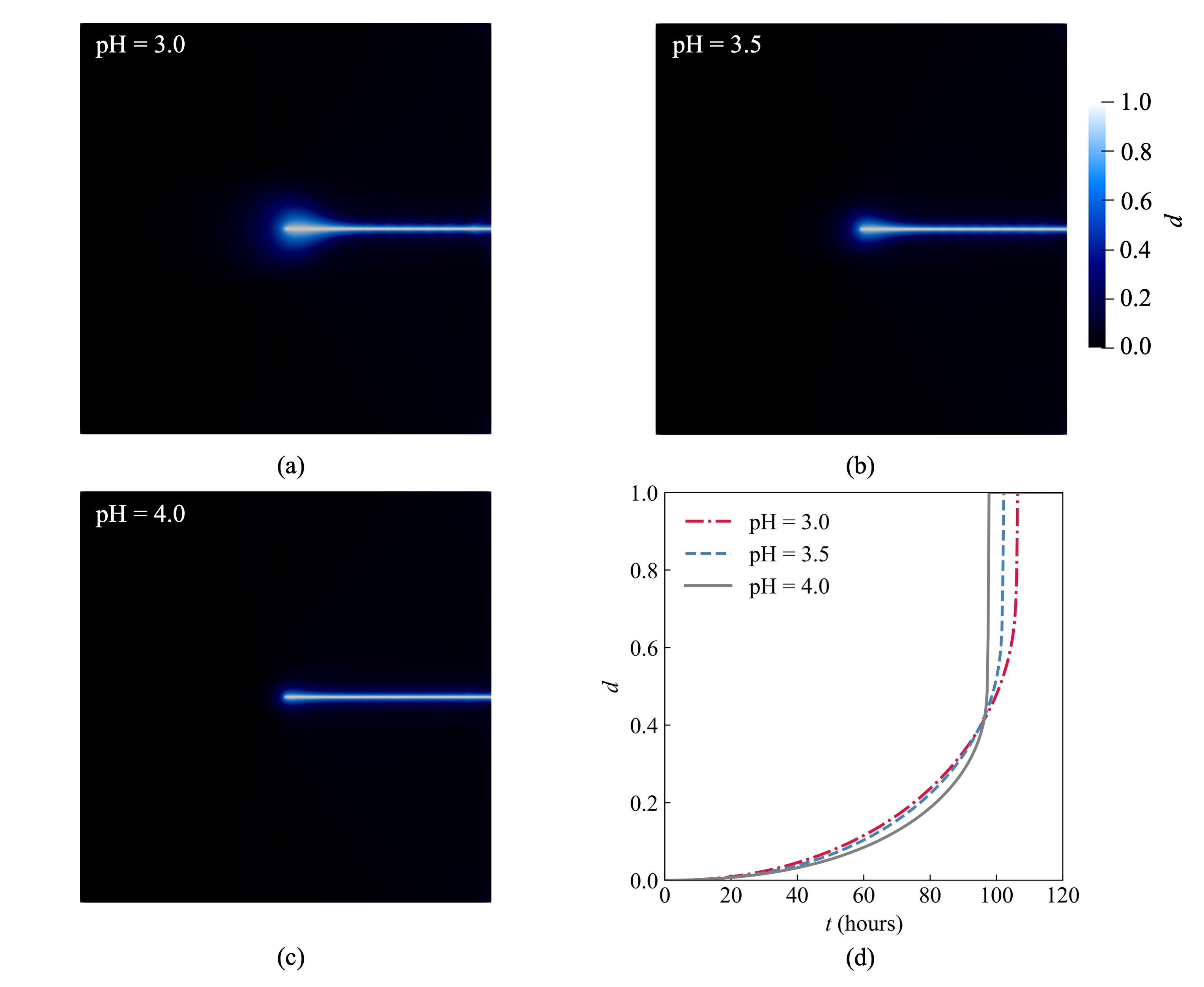}
    \caption{Distribution of phase-field damage at 120 h upon exposure to acidity levels of (a) pH = 3.0, (b) pH = 3.5, and (c) pH = 4.0. (d) Evolution of phase-field damage at the front of the crack tip in different chemical environments. The tracking point is marked in blue in Figure~\ref{fig:geometry}.}
    \label{fig:crack_ph}
\end{figure}

 Figures~\ref{fig:crack_ph}a-c display the distribution profile of the phase-field damage after 120 hours of exposure, under acidic environment of three different levels, with pH value of 3.0, 3.5, and 4.0, respectively. In the case of highest environmental intensity (i.e., pH = 3.0), the resulting process zone around the crack-tip appears to be the largest in size, among the three cases. Conversely, with an increased pH, i.e. the fracture is exposed to a less acidic environment, the size of the FPZ significantly shrinks. The evolution of phase-field damage over time at a point of 2 mm along the propagation direction in front of the original crack-tip, is plotted in Figure~\ref{fig:crack_ph}d. For all three environmental pH levels, the evolution of material property exhibits a more ductile behavior in comparison to the purely mechanical case shown in Figure~\ref{fig:comparison}a. It is also demonstrated in Figure~\ref{fig:crack_ph}d that as the environmental acidity increases (a decreasing pH value), the material response shifts towards a more ductile fracturing mode. Moreover, the critical crack onset, defined as the time required for the damage variable $d$ to reach 1.0, is delayed at lower pH values (as shown in Figure~\ref{fig:crack_ph}d), indicating that a more reactive environment exerts a ductilization effect on fracture propagation.

As addressed in Section~\ref{sec:length_coupling}, under chemo-mechanical feedback, fracture evolution is coupled with the reactive diffusion process of proton via the resulting mineral mass removal. On the one hand, the characteristics of the FPZ and the fracturing mode are tightly linked to the reactive environment to which the material is exposed to. On the other hand, the reaction-diffusion of proton is significantly affected by the micro-cracking process, as the newly generated crack walls create additional fluid pathways and surface areas available for chemical dissolution. The mutually promoting feedback between mechanical and chemical processes is successfully captured in our formulation. To illustrate this dynamic interplay during fracture propagation, Figures~\ref{fig:concentration_ph}a-c show the distribution of pH at 120 h after exposure to various chemical environments at constant pH value of 3.0, 3.5, and 4.0, respectively. Given that source point of proton release is at the crack surfaces, the lowest pH value is consistently observed at the crack opening, which increases gradually as it extends into the surrounding solid matrix. Meanwhile, as the fracture propagates forward, both the local diffusivity and the mineral dissolution in front of the propagating tip are enhanced as a result of damage evolution. Thereby, as seen in all the three profiles (Figures~\ref{fig:concentration_ph}a-c), the spatial distribution of acidity through the pH profile extends as the crack propagates, manifesting the extent of acid delivery into the solid matrix during fracturing.

\subsubsection{Acidity distribution with decreasing environmental pH}

\begin{figure}[t]
    \centering
    \includegraphics[width=0.8\textwidth]{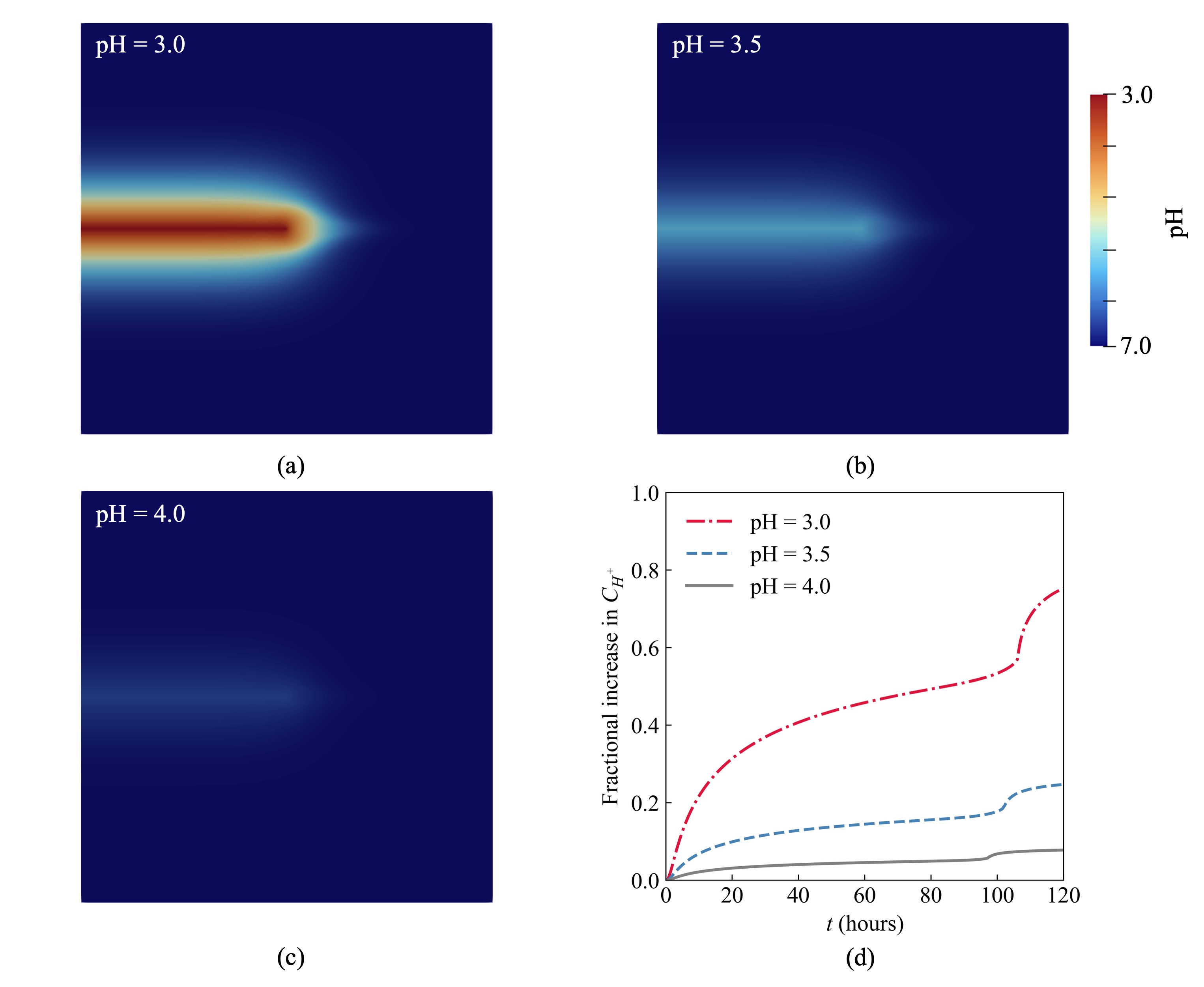}
    \caption{Distribution of pH at 120 h upon exposure to acidity levels of (a) pH = 3.0, (b) pH = 3.5, and (c) pH = 4.0. (d) Evolution of the fractional increase in acidity at the front of the crack tip in different chemical environments.}
    \label{fig:concentration_ph}
\end{figure}

Next, we define the fractional increase in local acidity as the change in proton concentration, $C_{\ce{H}^{+}}-C_n$, normalized by the maximum proton concentration corresponding to pH = 3.0. Here, $C_n$ denotes the initial proton concentration of pH = 7. This variable is therefore bounded between 0 and 1.0. Figure~\ref{fig:concentration_ph}d shows the evolution of the defined fractional increase in local acidity at the front of the crack-tip in the three prescribed chemical environments. Initially, the acidity increases rapidly as protons diffuse into the solid matrix from the source located at the crack surfaces. The rate of increase subsequently slows down, as a result of the decrease in the local concentration gradient. A distinct turning point is observed in each curve in Figure~\ref{fig:concentration_ph}d, corresponding to the onset of crack propagation. Beyond this point, the rise in local acidity experiences a significant acceleration, driven mainly by the enhanced diffusivity and mineral dissolution in front of the tip-point accompanied by the formation of the new crack, corresponding to $d$ approaching 1.0. Notably, with a higher environmental intensity (e.g. pH = 3.0), the critical turning point in proton concentration is slightly delayed compared with a more neutral environmental pH, which aligns with the damage evolution depicted in Figure~\ref{fig:crack_ph}d.

\subsubsection{Enhanced mineral dissolution with decreasing environmental pH}

\begin{figure}[t]
    \centering
    \includegraphics[width=0.8\textwidth]{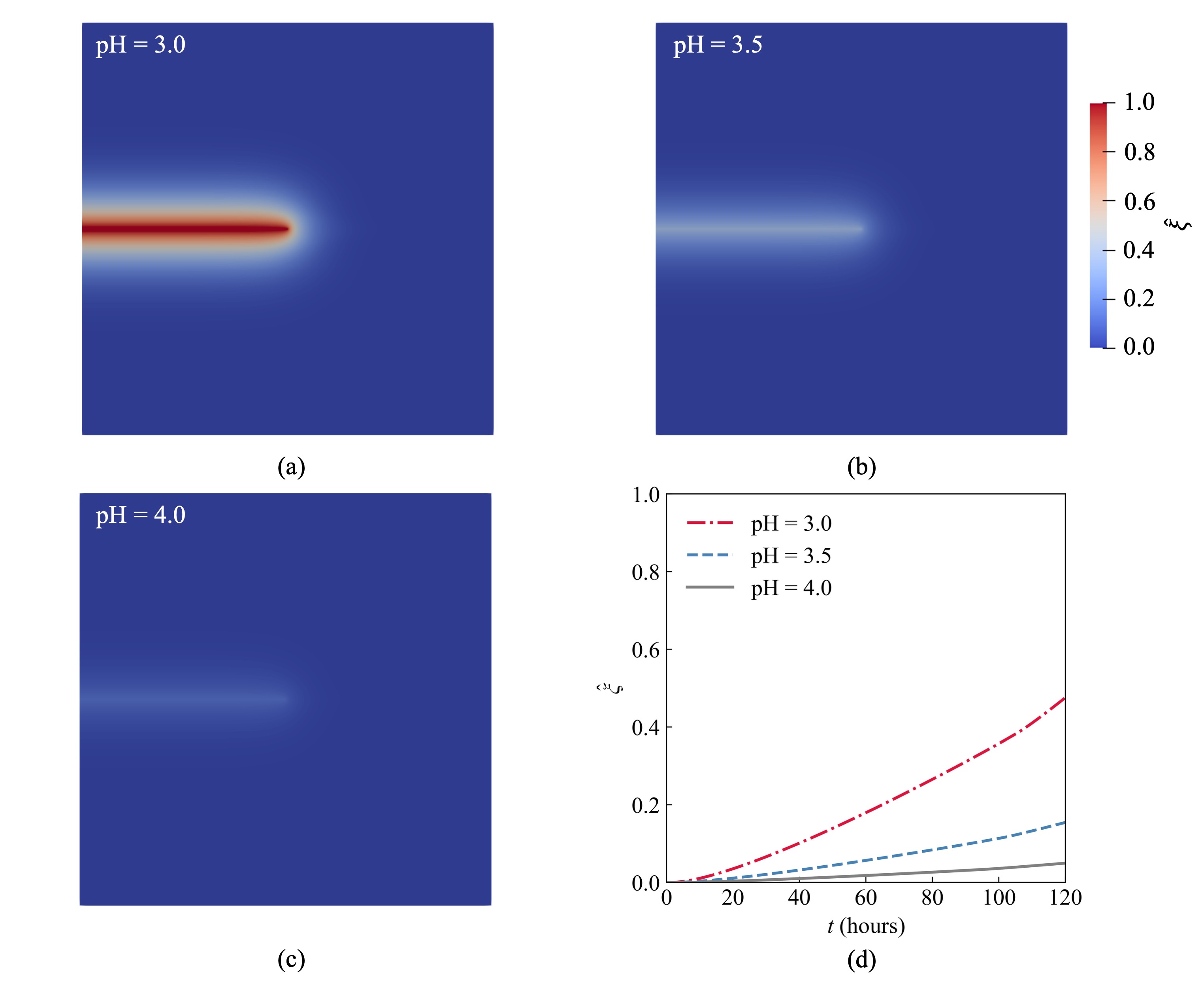}
    \caption{Distribution of normalized mass removal at 120 h upon exposure to acidity levels of (a) pH = 3.0, (b) pH = 3.5, and (c) pH = 4.0. (d) Evolution of normalized mass removal at the front of the crack tip in different chemical environments.}
    \label{fig:removal_ph}
\end{figure}

In our formulation, the crack length-scale is directly linked to the progress of chemical mass removal, which evolves according to the accumulated mineral dissolution triggered by the presence of acidic solutions. Figures~\ref{fig:removal_ph}a-c illustrate the spatial distribution of normalized mass removal after 120 hours of exposure to three prescribed environments with varying acid intensity. The normalized mass removal $\xi$ is defined as the ratio $\xi^{loc}/\xi^{loc}_{max}$, where $\xi^{loc}_{max}$ denotes the maximum value of local mass removal, which occurs in the highest acidity case (pH = 3.0). The simulation results show that as the environmental intensity increases, the extent of mass removal becomes more pronounced.  Notably, as shown in Figures~\ref{fig:removal_ph}a-c, after 120 hours, no significant mass removal is observed along the newly formed crack path. This observation is consistent with the constant crack width in the ``chemical-lag zone", where fracture propagation precedes chemical dissolution, as shown in Figure~\ref{fig:evolution}. Figure~\ref{fig:removal_ph}d provides a quantitative evaluation by plotting the temporal evolution of normalized mass removal in front of the crack tip under the three respective chemical environments. In all scenarios, the normalized mass removal increases over time, with the rate and total amount being highest in the most reactive environment (pH = 3.0). The environment with the lowest acidity (pH = 4.0) results in the least amount of accumulated mass removal. However, unlike the acidity evolution, the mass removal evolution in low-acidity cases does not exhibit a distinct acceleration point coinciding with the onset of crack propagation.
This occurs because chemical mass removal is inherently a time-cumulative process. For instance, under the condition of environmental pH = 4.0, the crack-induced enhancement of proton concentration is limited (Figure~\ref{fig:concentration_ph}d), resulting in a corresponding slow removal rate, with an insignificant acceleration in the mass removal evolution. In the case of environmental pH = 3.0, the chemical mass removal starts to accelerate upon the onset of crack propagation. In general, under an acidic environment, the accumulated mass removal is not expected to manifest as an abrupt surge at the onset of crack propagation.

\subsection{The effect of mechanical loading rate on crack propagation}
\label{sec:loading_rate_effect}

\subsubsection{Brittle-to-ductile transition with decreasing loading rate}
\begin{figure}[t]
    \centering
    \includegraphics[width=0.8\textwidth]{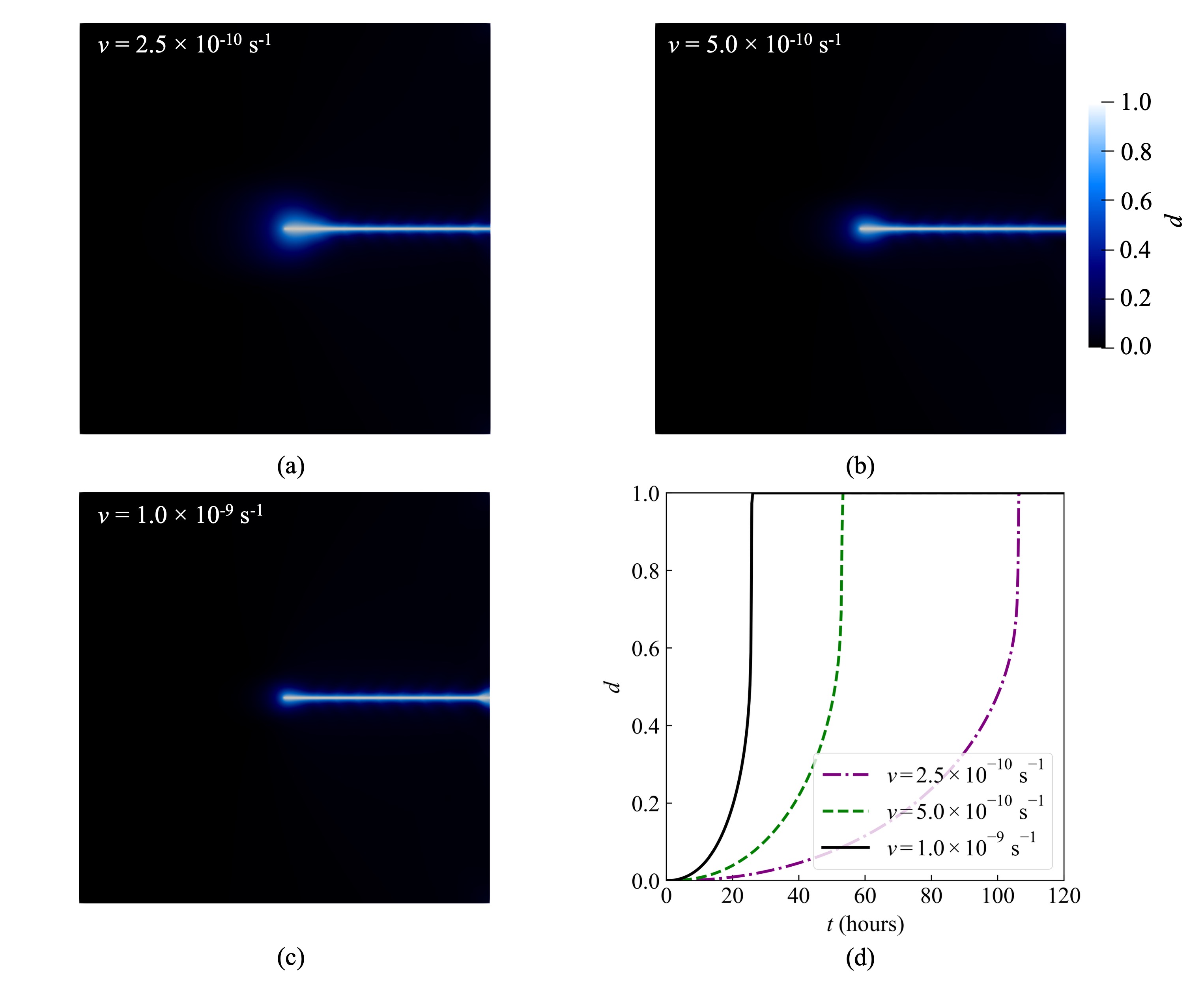}
    \caption{Distribution of phase-field damage at failure under tensile loading rates of (a) $v = 2.5 \times 10^{-10}$ s$^{-1}$, (b) $v = 5.0 \times 10^{-10}$ s$^{-1}$, and (c) $v = 1.0 \times 10^{-9}$ s$^{-1}$. (d) Evolution of phase-field damage at the front of the crack tip under different loading rates.}
    \label{fig:crack_rate}
\end{figure}

Next, we investigate the competition between mechanical and chemical effects on the onset of cracking, failure mode, and growth of the FPZ, by applying a variety of mechanical loading rates to the mode I fracture. Constant tensile extension rates of $v = 2.5 \times 10^{-10}$, $5.0 \times 10^{-10}$, and $1.0 \times 10^{-9}$ s$^{-1}$ were applied at the top and bottom boundaries, respectively. Figures~\ref{fig:crack_rate}a-c compare the phase-field damage distributions at the point of failure for these three cases, under the same environmental acidity fixed at pH = 3.0. In the base case with the slowest loading rate ($v = 2.5 \times 10^{-10}$ s$^{-1}$), catastrophic failure with the crack propagating through the sample occurs at approximately 110 hours. This relatively long duration allows for significant matrix dissolution around the initial crack tip, which in turn enlarges the FPZ. The size of the FPZ at failure is observed to decrease as the tensile loading rate increases, indicating that failure is dominated by mechanical effects at higher loading rates where chemical influence is insignificant throughout the process. Furthermore, Figure~\ref{fig:crack_rate}d illustrates the evolution of the damage $d$ at the front of the crack tip subject to varying mechanical loading rates, under identical chemical conditions. As anticipated, a higher loading rate leads to a faster onset of cracking. Under a rapid mechanical loading, the duration for chemical dissolution is limited, which results in a more brittle fracture behavior. Conversely, lower loading rates allow more time for chemically induced deterioration, causing the fracturing mode to transit towards a more ductile behavior, manifested as a more gradual increase in $d$ over time.

\subsubsection{Acidity distribution with decreasing loading rate}
\begin{figure}[t]
    \centering
    \includegraphics[width=0.8\textwidth]{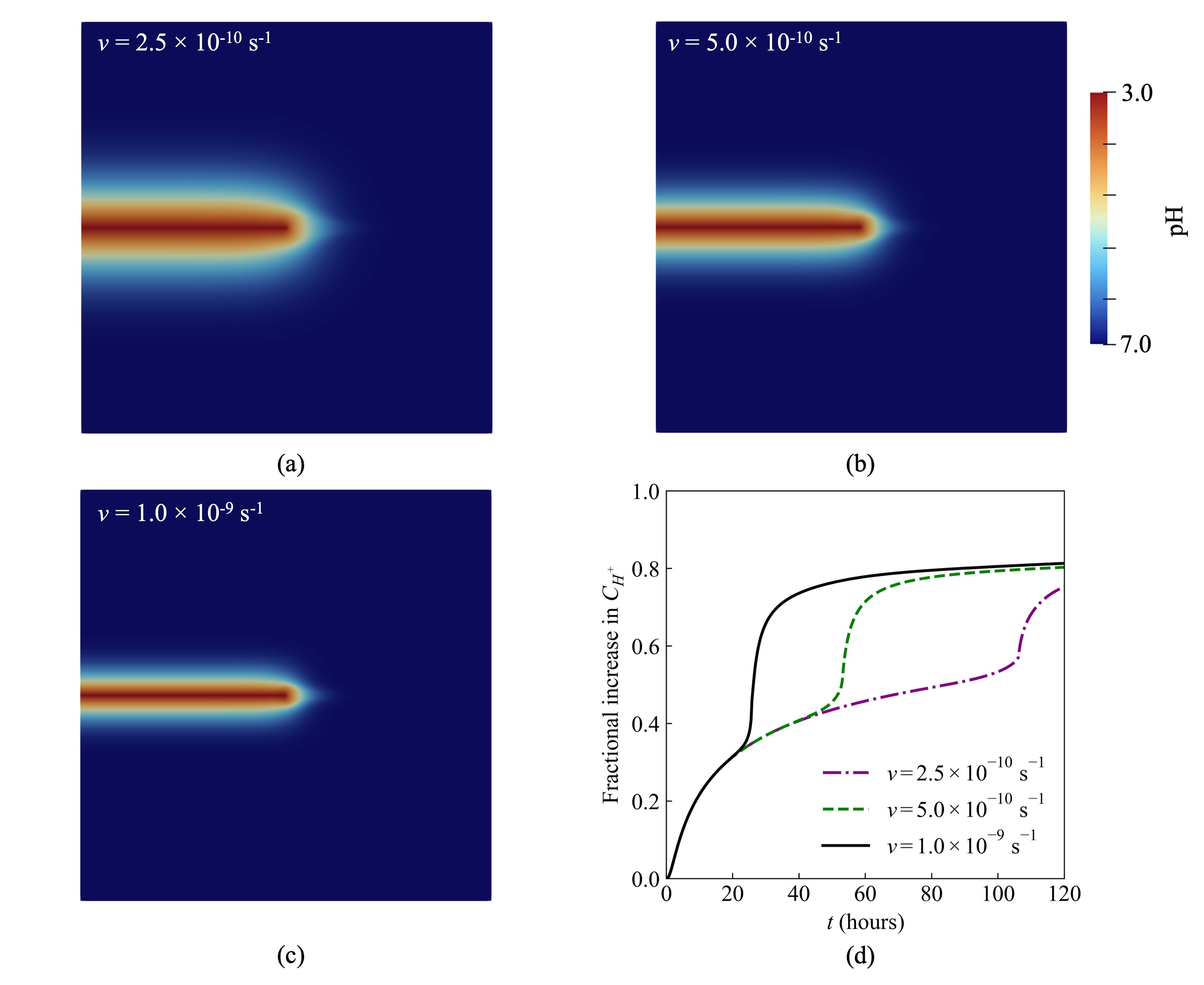}
    \caption{Distribution of pH at failure under tensile loading rates of (a) $v = 2.5 \times 10^{-10}$ s$^{-1}$, (b) $v = 5.0 \times 10^{-10}$ s$^{-1}$, and (c) $v = 1.0 \times 10^{-9}$ s$^{-1}$. (d) Evolution of the fractional increase in acidity at the front of the crack tip under different loading rates.}
    \label{fig:concentration_rate}
\end{figure}

Figures~\ref{fig:concentration_rate}a-c plot the pH distribution within the samples at the time of material failure for different tensile loading rates, when exposed to an identical environmental pH of 3.0. The most diffusive proton distribution profile at failure is observed for the slowest mechanical loading rate. In contrast, under rapid mechanical loading conditions, the proton distribution profile is substantially less diffusive, suggesting a relatively limited acid delivery. Notably, in all three cases the acid delivery ahead of the crack-tip along the propagation path appears to be significant, which results from an enhanced damage development under chemo-mechanical feedback. Figure~\ref{fig:concentration_rate}d provides a quantitative comparison by plotting the time evolution of the fractional increase in acidity in front of the crack tip. Before the onset of crack propagation, the acidity evolution experiences an identical growth for all the three cases because, in the absence of significant damage at the front of the tip, the reactive diffusion of protons remains independent of the mechanical loading rate. Upon the onset of crack propagation, a corresponding surge in the fractional increase in proton concentration exhibits. As demonstrated in Figure~\ref{fig:concentration_rate}d, the slowest mechanical loading rate leads to the most postponed failure, as well as the smallest final fractional increase in proton concentration at the crack tip due to the relatively limited pathways for acid delivery.

\subsubsection{Enhanced mineral dissolution with decreasing loading rate}
\begin{figure}[t]
    \centering
    \includegraphics[width=0.8\textwidth]{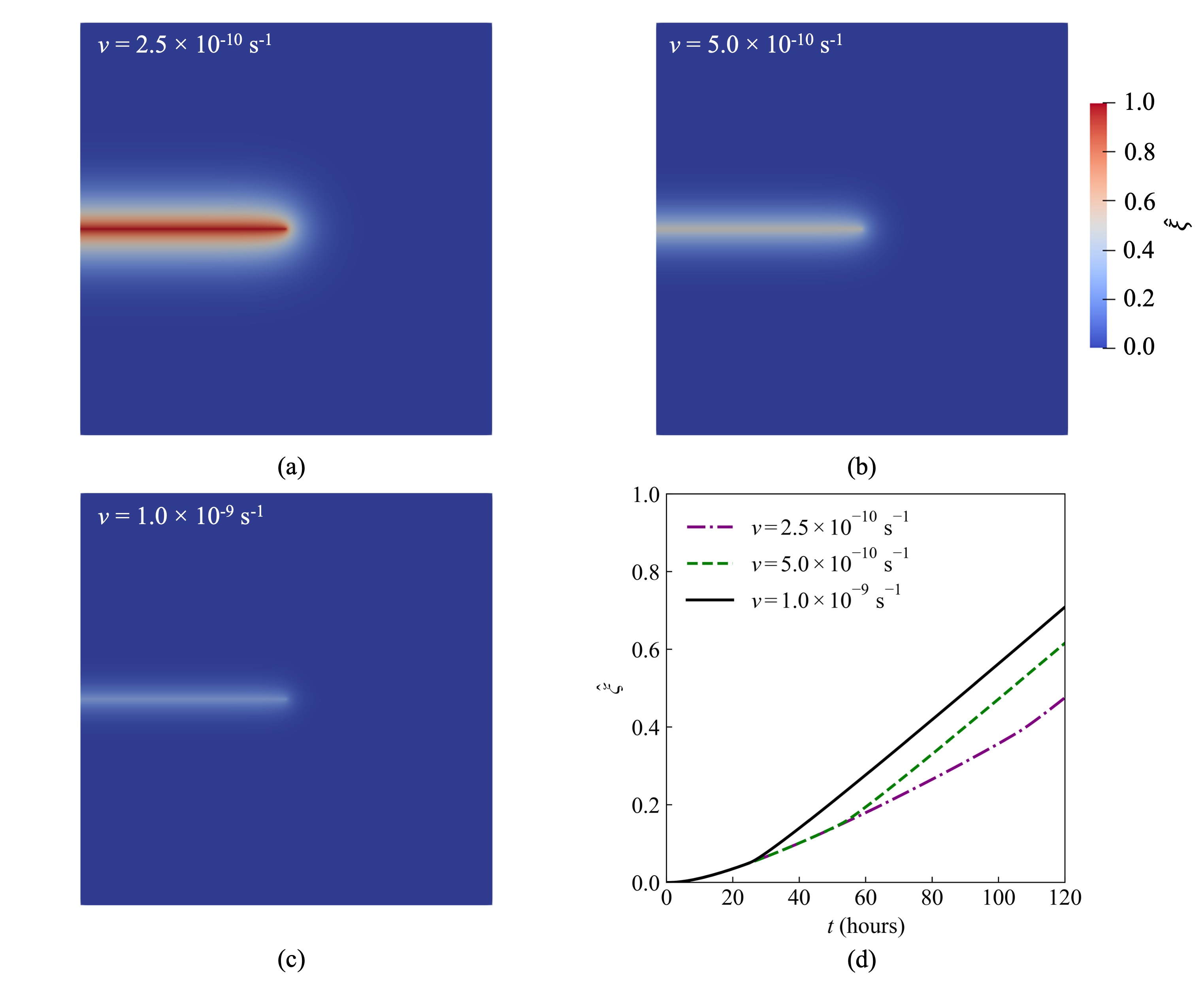}
    \caption{Distribution of normalized mass removal at failure under tensile loading rates of (a) $v = 2.5 \times 10^{-10}$ s$^{-1}$, (b) $v = 5.0 \times 10^{-10}$ s$^{-1}$, and (c) $v = 1.0 \times 10^{-9}$ s$^{-1}$. (d) Evolution of normalized mass removal at the front of the crack tip under different loading rates.}
    \label{fig:removal_rate}
\end{figure}

The distribution of normalized mass removal at failure, shown in Figures~\ref{fig:removal_rate}a-c, exhibits a similar dependence on the tensile loading rate. As the loading rate increases, the onset of crack propagation accelerates, allowing less time for the acidic agents to diffuse and dissolve the mineral matrix before failure. As a result, for the fastest loading rate ($v = 1.0 \times 10^{-9}$ s$^{-1}$), the extent of mass removal at failure is minimal. The lack of chemical deterioration supports the observation of the smaller FPZ for this case shown in Figure~\ref{fig:crack_rate}c. Figure~\ref{fig:removal_rate}d presents the evolution of normalized mass removal at a point in front of the crack tip for each loading rate. Initially, the mass removal evolves identically in all cases. However, as the loading rate decreases, the critical acceleration point in mass removal is progressively delayed. Note that the acceleration here does not coincide exactly with the damage variable $d$ approaching 1, neither the surge in local proton concentration. There is a noticeable time lag due to the fact that mass removal is a cumulative, time-dependent process. For the slowest mechanical loading rate ($v = 2.5 \times 10^{-10}$ s$^{-1}$), this acceleration phase occurs much later. Consequently, the total accumulated mass removal at 120 hours is lower than the counterpart in the other two cases.

\subsection{Stress response at the front of the crack tip}
\label{sec:stress_resp}

\begin{figure}[t]
    \centering
    \includegraphics[width=0.85\textwidth]{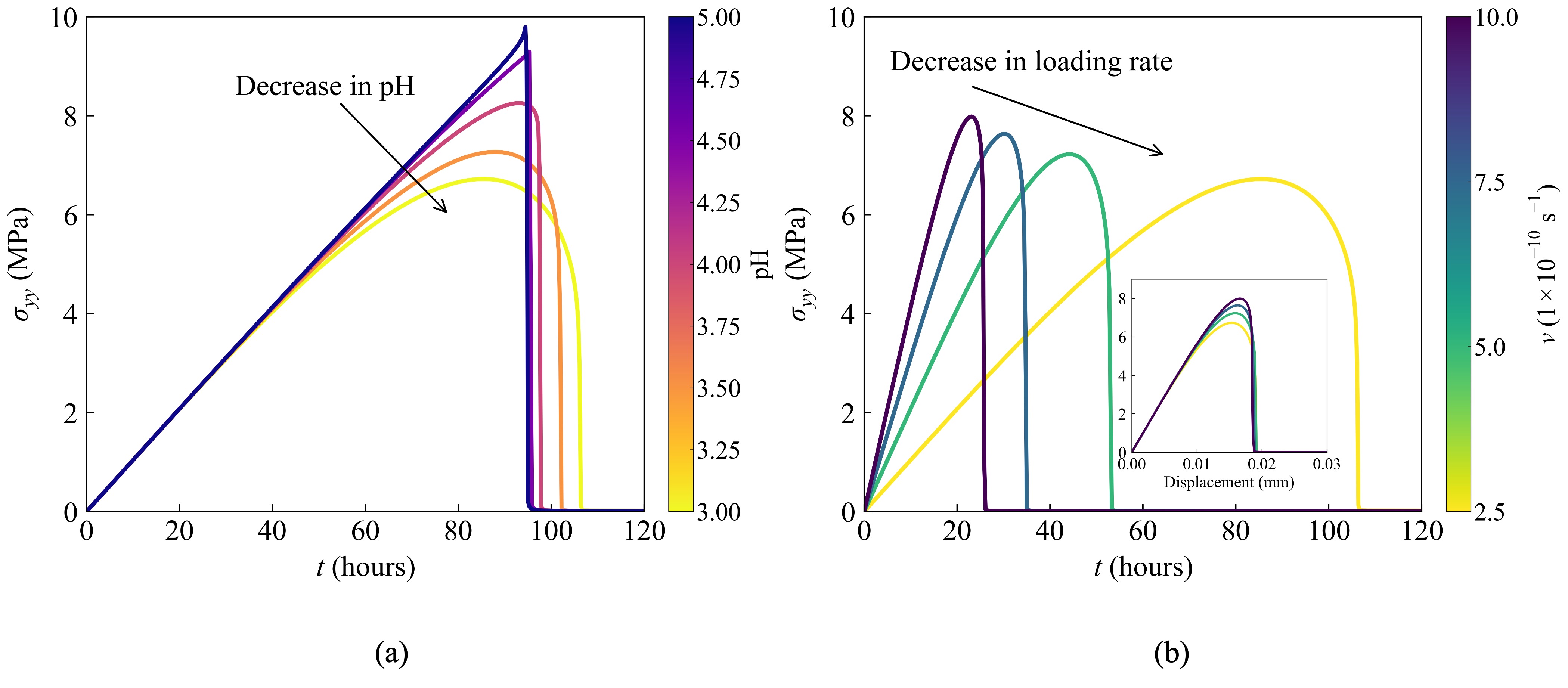}
    \caption{Temporal evolution of the circumferential stress at the front of the crack tip of the pre-notch sample: (a) considering various acidic environments, and (b) considering different mechanical loading rates. The inset in (b) shows the same stress evolution against the displacement.}
    \label{fig:stress}
\end{figure}

Figure~\ref{fig:stress}a shows the temporal evolution of the circumferential stress ($\sigma_{yy}$) 2 mm ahead of the initial crack tip. Under an identical displacement-controlled loading condition, the local stress builds up gradually, reaching a peak value before dropping abruptly as the fracture propagates through the sample. In the initial stage (approximately 0--40 hours), the material stiffness remains constant in all cases due to negligible dissolution. However, as chemical and mechanical damage accumulate, a discernible reduction in stiffness shows up, prior to failure. This stiffness degradation is notably more pronounced in highly acidic environments (lower pH), resulting in a reduced peak circumferential stress. As indicated by the arrow in Figure~\ref{fig:stress}a, the alleviation of the stress singularity at the crack tip is attributed to the chemically induced expansion of the FPZ, which diffuses the stress concentration. By applying an array of different environmental pH values ranging from 3.0 to 4.0, we reveal a clear brittle-to-ductile transition, characterized by improved compliance as acidity increases. As discussed previously in Section~\ref{sec:pH_effect}, the delay of fracture onset under low pH conditions (Figure~\ref{fig:crack_ph}) is also confirmed here by the delayed drop in the corresponding stress field. Furthermore, our model regularizes the singular stress field predicted by linear elastic fracture mechanics (LEFM) theory through capturing the chemically enhanced growth of the FPZ at the crack tip.

In the same vein, Figure~\ref{fig:stress}b presents the evolution of circumferential stress under varying mechanical loading rates under a fixed environmental pH of 3.0 (baseline case). With a decrease in loading rate and the corresponding increase in chemical exposure time, the maximum stress in the crack opening direction is lowered, suggesting a strength degradation caused by cumulative mineral mass removal (see Figure~\ref{fig:removal_rate}). Moreover, the post-peak response shows that prolonged acidic exposure induces a ductilization of the failure mode, consistent with the findings in Section~\ref{sec:loading_rate_effect}. The inset plots the stress evolution against displacement, demonstrating the stress responses on a comparable scale despite of the different loading rates. The stress response in Figure~\ref{fig:stress}b also exhibits an apparent shift from a brittle mode towards a more ductile behavior, with the decrease in the loading rate. 

\subsection{Comparisons of the width of FPZ}
\label{sec:width_fpz}

\begin{figure}[t]
    \centering
    \includegraphics[width=0.85\textwidth]{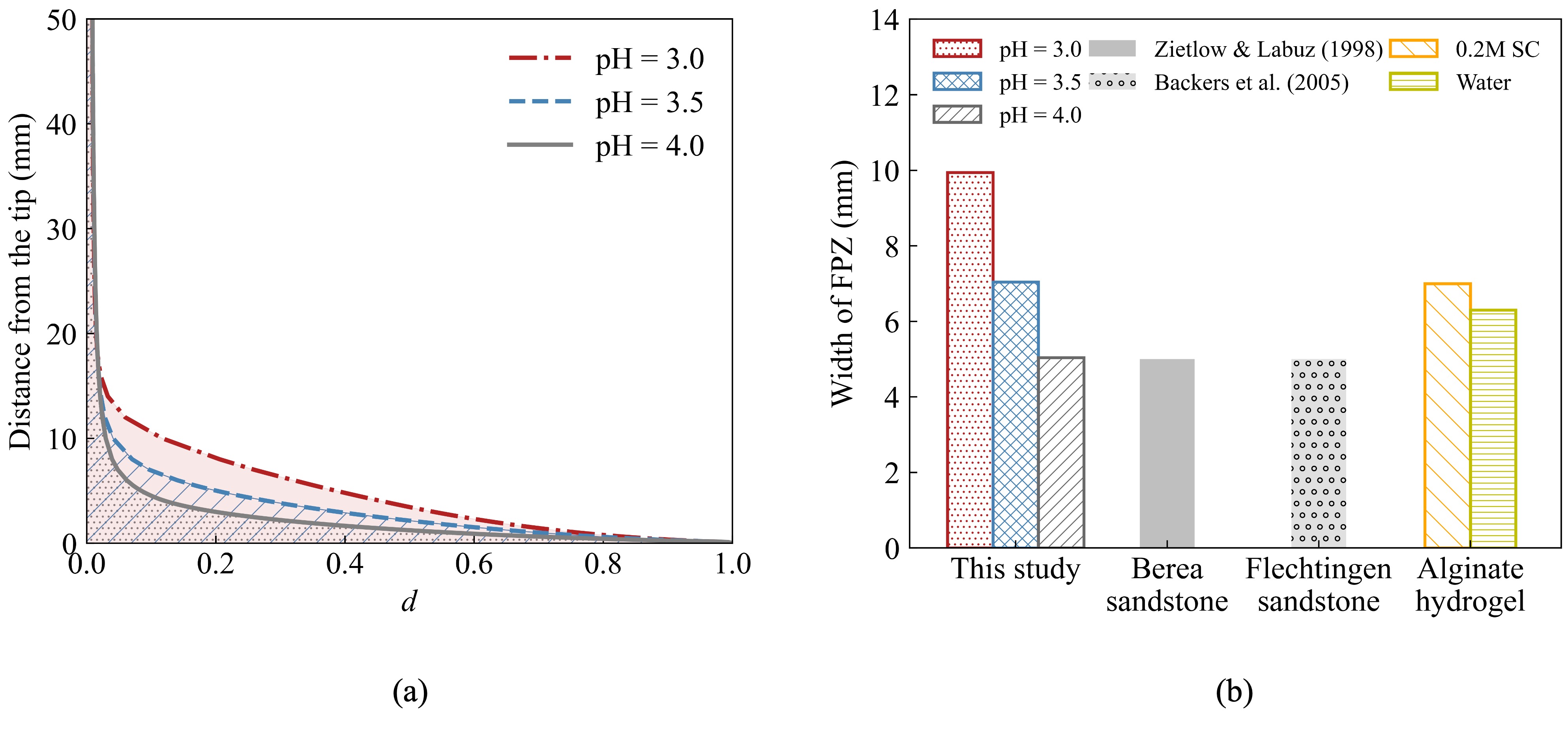}
    \caption{(a) Determination of the width of the FPZ on a diffusive damage profile. (b) Comparison of the width of FPZ predicted by this study and measured via different experimental techniques \citep{zietlow_measurement_1998,backers_tensile_2005,chen_single_2024,chen_chemically_2025}.}
    \label{fig:fpz}
\end{figure}

Using the same setup as in Section~\ref{sec:pH_effect}, we investigate the width of FPZ in different reactive environments. To obtain the spatial extent of diffusive damage, we define a characteristic width of the FPZ, denoted as $S$. The width is evaluated along the direction perpendicular to fracture propagation. As shown in Figure~\ref{fig:fpz}a, under the symmetric setting, the total effective width is determined by integrating the diffusive phase-field damage profile over the half-domain:
\begin{equation}
S = 2 \int_{0}^{X_{max}} d(x) \mathrm{d} x \,,
\end{equation}
where $X_{max}$ represents the boundary of the half-domain. The formulation provides an equivalent length scale that captures the cumulative intensity of the diffusive damage field. This metric was measured at the end of each simulation. The calculated width of FPZ ranges from 5.0 to 9.9 mm, decreasing as the environmental pH value increases.

Figure~\ref{fig:fpz}b compares the widths calculated in this study against experimental data obtained using various techniques. For natural rock materials, Acoustic Emission (AE) monitoring is widely used to track the formation of microcracks. It offers a crucial approach for validating the calculated characteristic width in this work. For instance, \cite{zietlow_measurement_1998} performed three-point bending tests on Berea sandstone beams and determined that the intrinsic width of FPZ is a material property measuring approximately 5.0 mm through AE monitoring. Similarly, \cite{backers_tensile_2005} investigated Mode I fracture in Flechtingen sandstone using chevron-bend tests, confirming that the FPZ width perpendicular to the fracture plane remained constant at about 5.0 mm, independent of different applied loading rates.

Expanding the validation to other brittle solids, \cite{chen_single_2024,chen_fluid_2026} employed a Hele-Shaw cell configuration to investigate fluid-driven Mode I fracture in alginate hydrogel, a transparent analogue material for low-permeability brittle rocks. By injecting fluorescent-dyed fluids, the experiment successfully visualized the infiltration zone surrounding the propagating crack tip, where micro damage was generated, thereby identifying a characteristic length scale of the FPZ. Comparisons between injecting a chemically reactive fluid and a non-reactive fluid reveal that dissolution can significantly enhance the size of the damaged zone. The width of FPZ was measured to be 6.3 mm when subject to deionized water injection and increased to 7.0 mm under the injection of reactive 0.2M sodium citrate (SC). The laboratory observation that stronger chemical interactions enlarge the damage zone aligns with the results of the present simulation, i.e. increased acidity-induced mass removal enhances the width of FPZ. To summarize, these comparisons across natural rocks and hydrogel analogues show excellent agreement with the characteristic width of FPZ predicted by our chemo-mechanical phase-field simulations, as demonstrated in Figure~\ref{fig:fpz}b.

\section{Conclusion}
This work presents a chemo-mechanical phase-field model for elucidating the complex interplay between chemical mass removal and fracture propagation in dissolvable rocks subject to reactive environments. A key novelty of this approach lies in its intrinsic coupling scheme between the progress of mineral dissolution and rock deformation, represented by one single phase-field variable. The size of the process zone in front of the propagating crack-tip is hence defined by the reaction-diffusion process depending on the damage evolution. By capturing the two-way coupled chemo-mechanical feedback, crack propagation under various chemical intensity and mechanical loading conditions is investigated using the developed model. The main contributions are summarized as follows:
\begin{enumerate}[(1)]
    \item The coupled framework effectively captures the mutual feedback between chemical mass removal and mechanical damage, distinguishing reactive cracking from purely mechanically driven fracture. One particular feature is the development of an enlarged and diffusive FPZ at the crack tip. The chemically induced widening of the FPZ blunts the sharp crack tip, resulting in a distinct ductilization effect characterized by a more gradual accumulation of damage and a delayed onset of macroscopic failure.

    \item The transition between brittle and ductile failure mode is determined by the competing timescales of mechanical loading and chemical reaction. Our results indicate that highly reactive environments (low pH) enhance matrix dissolution surrounding the crack tip and promote ductile fracture behavior. In contrast, rapid mechanical loading suppresses chemical effects by limiting the interaction duration, preserving the brittle nature of the fracture. Consequently, the failure mode is determined by the balance between the rate of external loading and the rate of chemical degradation.

    \item We demonstrated that the chemically induced expansion of the FPZ effectively alleviates the stress concentration near the crack tip. Exposure to reactive environments results in a pronounced degradation in stiffness and a marked reduction in the circumferential stress in front of the crack-tip before material failure. The local stress evolution supports the macroscopic observation of the brittle-to-ductile transition, where cumulative mass removal enhances the matrix compliance and suppresses brittle fracture behavior.

\end{enumerate}

Beyond single crack propagation, the proposed coupling between chemical mass removal and fracture length scale also provides a basis for our investigation into spatially organized crack patterns. In our companion study \citep{wu_companion}, this micromechanical linkage is applied to explain the formation of regularly spaced parallel cracks under chemo-mechanical conditions.

\section{Declaration of Competing Interest}
The authors declare that they have no known competing financial interests or personal relationships that could have appeared to influence the work reported in this paper.

\section{Data availability}
Data will be made available upon reasonable request.

\section{Declaration of Generative AI and AI-assisted technologies in the writing process}
During the preparation of this work, the authors used ChatGPT to improve the readability and language of the manuscript. After using this tool, the authors reviewed and edited the content as needed and take full responsibility for the content of the published article.

\section{Acknowledgements}
We acknowledge the support from the National Natural Science Foundation of China (NSFC 12302235), the Research Grants Council of Hong Kong (GRF 17206521), and the U.S. National Science Foundation (NSF Projects CMMI-2042325 and 2332069).

\bibliographystyle{elsarticle-harv} 
\bibliography{refs}

@article{miehe_phase_2010,
	title = {A phase field model for rate-independent crack propagation: {Robust} algorithmic implementation based on operator splits},
	volume = {199},
	issn = {0045-7825},
	shorttitle = {A phase field model for rate-independent crack propagation},
	doi = {10.1016/j.cma.2010.04.011},
	number = {45},
	journal = {Computer Methods in Applied Mechanics and Engineering},
	author = {Miehe, Christian and Hofacker, Martina and Welschinger, Fabian},
	year = {2010},
	pages = {2765--2778},
}

@article{miehe_thermodynamically_2010-1,
	title = {Thermodynamically consistent phase-field models of fracture: {Variational} principles and multi-field {FE} implementations},
	volume = {83},
	issn = {1097-0207},
	shorttitle = {Thermodynamically consistent phase-field models of fracture},
	doi = {10.1002/nme.2861},
	number = {10},
	journal = {International Journal for Numerical Methods in Engineering},
	author = {Miehe, C. and Welschinger, F. and Hofacker, M.},
	year = {2010},
	pages = {1273--1311},
}

@article{bilgen_crack-driving_2019,
	title = {On the crack-driving force of phase-field models in linearized and finite elasticity},
	volume = {353},
	issn = {0045-7825},
	doi = {10.1016/j.cma.2019.05.009},
	journal = {Computer Methods in Applied Mechanics and Engineering},
	author = {Bilgen, Carola and Weinberg, Kerstin},
	year = {2019},
	pages = {348--372},
}

@article{ip_modeling_2023,
	title = {Modeling heterogeneity and permeability evolution in a compaction band using a phase-field approach},
	volume = {181},
	issn = {0022-5096},
	doi = {10.1016/j.jmps.2023.105441},
	journal = {Journal of the Mechanics and Physics of Solids},
	author = {Ip, Sabrina C. Y. and Borja, Ronaldo I.},
	year = {2023},
	pages = {105441},
}

@article{dugdale_yielding_1960,
	title = {Yielding of steel sheets containing slits},
	volume = {8},
	issn = {0022-5096},
	doi = {10.1016/0022-5096(60)90013-2},
	number = {2},
	journal = {Journal of the Mechanics and Physics of Solids},
	author = {Dugdale, D. S.},
	year = {1960},
	pages = {100--104},
}

@article{belytschko_elastic_1999,
	title = {Elastic crack growth in finite elements with minimal remeshing},
	volume = {45},
	copyright = {Copyright © 1999 John Wiley \& Sons, Ltd.},
	issn = {1097-0207},
	doi = {10.1002/(SICI)1097-0207(19990620)45:5<601::AID-NME598>3.0.CO;2-S},
	number = {5},
	journal = {International Journal for Numerical Methods in Engineering},
	author = {Belytschko, T. and Black, T.},
	year = {1999},
	pages = {601--620},
}

@article{moes_finite_1999,
	title = {A finite element method for crack growth without remeshing},
	volume = {46},
	copyright = {Copyright © 1999 John Wiley \& Sons, Ltd.},
	issn = {1097-0207},
	doi = {10.1002/(SICI)1097-0207(19990910)46:1<131::AID-NME726>3.0.CO;2-J},
	number = {1},
	journal = {International Journal for Numerical Methods in Engineering},
	author = {Moës, Nicolas and Dolbow, John and Belytschko, Ted},
	year = {1999},
	pages = {131--150},
}

@article{francfort_revisiting_1998,
	title = {Revisiting brittle fracture as an energy minimization problem},
	volume = {46},
	issn = {0022-5096},
	doi = {10.1016/S0022-5096(98)00034-9},
	number = {8},
	journal = {Journal of the Mechanics and Physics of Solids},
	author = {Francfort, G. A. and Marigo, J. -J.},
	year = {1998},
	pages = {1319--1342},
}

@article{bourdin_numerical_2000,
	title = {Numerical experiments in revisited brittle fracture},
	volume = {48},
	issn = {0022-5096},
	doi = {10.1016/S0022-5096(99)00028-9},
	number = {4},
	journal = {Journal of the Mechanics and Physics of Solids},
	author = {Bourdin, B. and Francfort, G. A. and Marigo, J-J.},
	year = {2000},
	pages = {797--826},
}

@misc{Raccoon2020,
  author = {Hu, T.},
  title  = {{RACCOON} - {A} parallel finite-element code specialized in phase-field for fracture},
  howpublished = "\url{https://hugary1995.github.io/raccoon/index.html}",
  year = {2020}
}

@article{permann_moose_2020,
	title = {{MOOSE}: {Enabling} massively parallel multiphysics simulation},
	volume = {11},
	issn = {2352-7110},
	shorttitle = {{MOOSE}},
	doi = {10.1016/j.softx.2020.100430},
	journal = {SoftwareX},
	author = {Permann, Cody J. and Gaston, Derek R. and Andrš, David and Carlsen, Robert W. and Kong, Fande and Lindsay, Alexander D. and Miller, Jason M. and Peterson, John W. and Slaughter, Andrew E. and Stogner, Roy H. and Martineau, Richard C.},
	year = {2020},
	pages = {100430},
}

@article{schuler_chemo-mechanical_2020,
	title = {Chemo-mechanical phase-field modeling of dissolution-assisted fracture},
	volume = {362},
	issn = {0045-7825},
	doi = {10.1016/j.cma.2020.112838},
	journal = {Computer Methods in Applied Mechanics and Engineering},
	author = {Schuler, Louis and Ilgen, Anastasia G. and Newell, Pania},
	year = {2020},
	pages = {112838},
}

@article{guo_reactive-transport_2024,
	title = {A reactive-transport phase-field modelling approach of chemo-assisted cracking in saturated sandstone},
	volume = {419},
	issn = {0045-7825},
	doi = {10.1016/j.cma.2023.116645},
	journal = {Computer Methods in Applied Mechanics and Engineering},
	author = {Guo, Yongfan and Na, SeonHong},
	year = {2024},
	pages = {116645},
}

@article{hu_modeling_2019,
	title = {Modeling of subcritical cracking in acidized carbonate rocks via coupled chemo-elasticity},
	volume = {19},
	issn = {2352-3808},
	doi = {10.1016/j.gete.2019.01.003},
	journal = {Geomechanics for Energy and the Environment},
	author = {Hu, ManMan and Hueckel, Tomasz},
	year = {2019},
	pages = {100114},
}

@article{tang_reactive_2023,
	title = {A {Reactive}-{Chemo}-{Mechanical} {Model} for {Weak} {Acid}-{Assisted} {Cavity} {Expansion} in {Carbonate} {Rocks}},
	volume = {56},
	issn = {1434-453X},
	doi = {10.1007/s00603-022-03077-2},
	number = {1},
	journal = {Rock Mechanics and Rock Engineering},
	author = {Tang, XiaoJie and Hu, ManMan},
	year = {2023},
	pages = {515--533},
}

@article{tang_acid-assisted_2024,
	title = {Acid-assisted subcritical blunt-tip crack propagation in carbonate rocks},
	volume = {19},
	issn = {1861-1133},
	doi = {10.1007/s11440-024-02288-x},
	number = {5},
	journal = {Acta Geotechnica},
	author = {Tang, XiaoJie and Hu, ManMan},
	year = {2024},
	pages = {3095--3113},
}

@article{tang_effect_2025,
	title = {The {Effect} of {Chemical} {Environment} on {Crack} {Propagation} in {Pressure}-{Sensitive} {Rocks}},
	issn = {1434-453X},
	doi = {10.1007/s00603-025-04485-w},
	journal = {Rock Mechanics and Rock Engineering},
	author = {Tang, XiaoJie and Hu, ManMan},
	year = {2025},
}

@article{hu_environmentally_2013,
	title = {Environmentally enhanced crack propagation in a chemically degrading isotropic shale},
	volume = {63},
	issn = {0016-8505},
	doi = {10.1680/geot.SIP13.P.020},
	number = {4},
	journal = {Géotechnique},
	author = {Hu, M.m. and Hueckel, T.},
	year = {2013},
	pages = {313--321},
}

@article{ling_probing_2022,
	title = {Probing multiscale dissolution dynamics in natural rocks through microfluidics and compositional analysis},
	volume = {119},
	doi = {10.1073/pnas.2122520119},
	number = {32},
	journal = {Proceedings of the National Academy of Sciences},
	author = {Ling, Bowen and Sodwatana, Mo and Kohli, Arjun and Ross, Cynthia M. and Jew, Adam and Kovscek, Anthony R. and Battiato, Ilenia},
	year = {2022},
	pages = {e2122520119},
}

@article{borja_constitutive_2023,
	title = {A constitutive framework for rocks undergoing solid dissolution},
	volume = {173},
	issn = {0022-5096},
	doi = {10.1016/j.jmps.2023.105198},
	journal = {Journal of the Mechanics and Physics of Solids},
	author = {Borja, Ronaldo I. and Chen, Wei and Odufisan, Alesanmi R.},
	year = {2023},
	pages = {105198},
}

@article{mollaali_variational_2025,
	title = {Variational {Phase}-{Field} {Fracture} {Approach} in {Reactive} {Porous} {Media}},
	volume = {126},
	issn = {1097-0207},
	doi = {10.1002/nme.7621},
	number = {1},
	journal = {International Journal for Numerical Methods in Engineering},
	author = {Mollaali, Mostafa and Yoshioka, Keita and Lu, Renchao and Montoya, Vanessa and Vilarrasa, Victor and Kolditz, Olaf},
	year = {2025},
	pages = {e7621},
}

@article{wu_phase-field_2016,
	series = {Phase {Field} {Approaches} to {Fracture}},
	title = {A phase-field approach to fracture coupled with diffusion},
	volume = {312},
	issn = {0045-7825},
	doi = {10.1016/j.cma.2016.05.024},
	journal = {Computer Methods in Applied Mechanics and Engineering},
	author = {Wu, T. and De Lorenzis, L.},
	year = {2016},
	pages = {196--223},
}

@article{wu_onset_2025,
	title = {Onset of reactive brittle cracking in sandstones: {DEM}-informed phase-field modeling},
	volume = {196},
	issn = {1365-1609},
	shorttitle = {Onset of reactive brittle cracking in sandstones},
	doi = {10.1016/j.ijrmms.2025.106319},
	journal = {International Journal of Rock Mechanics and Mining Sciences},
	author = {Wu, Fanyu and Sac-Morane, Alexandre and Rattez, Hadrien and Veveakis, Manolis and Hu, Manman},
	year = {2025},
	pages = {106319},
}

@article{guevel_darcy_2023,
	title = {A {Darcy}–{Cahn}–{Hilliard} model of multiphase fluid-driven fracture},
	volume = {181},
	issn = {0022-5096},
	doi = {10.1016/j.jmps.2023.105427},
	journal = {Journal of the Mechanics and Physics of Solids},
	author = {Guével, Alexandre and Meng, Yue and Peco, Christian and Juanes, Ruben and Dolbow, John E.},
	year = {2023},
	pages = {105427},
}

@article{fei_phase-field_2023,
	title = {A phase-field model for hydraulic fracture nucleation and propagation in porous media},
	volume = {47},
	copyright = {© 2023 John Wiley \& Sons Ltd.},
	issn = {1096-9853},
	doi = {10.1002/nag.3612},
	number = {16},
	journal = {International Journal for Numerical and Analytical Methods in Geomechanics},
	author = {Fei, Fan and Costa, Andre and Dolbow, John E. and Settgast, Randolph R. and Cusini, Matteo},
	year = {2023},
	pages = {3065--3089},
}

@article{chukwudozie_variational_2019,
	title = {A variational phase-field model for hydraulic fracturing in porous media},
	volume = {347},
	issn = {0045-7825},
	doi = {10.1016/j.cma.2018.12.037},
	journal = {Computer Methods in Applied Mechanics and Engineering},
	author = {Chukwudozie, Chukwudi and Bourdin, Blaise and Yoshioka, Keita},
	year = {2019},
	pages = {957--982},
}

@article{cui_phase_2021,
	title = {A phase field formulation for dissolution-driven stress corrosion cracking},
	volume = {147},
	issn = {0022-5096},
	doi = {10.1016/j.jmps.2020.104254},
	journal = {Journal of the Mechanics and Physics of Solids},
	author = {Cui, Chuanjie and Ma, Rujin and Martínez-Pañeda, Emilio},
	year = {2021},
	pages = {104254},
}

@article{martinez-paneda_phase_2018,
	title = {A phase field formulation for hydrogen assisted cracking},
	volume = {342},
	issn = {0045-7825},
	doi = {10.1016/j.cma.2018.07.021},
	journal = {Computer Methods in Applied Mechanics and Engineering},
	author = {Martínez-Pañeda, Emilio and Golahmar, Alireza and Niordson, Christian F.},
	year = {2018},
	pages = {742--761},
}

@article{mishra_fracture_2026,
	title = {Fracture in concrete: {X}-ray tomography with in-situ testing, digital volume correlation and phase-field modeling},
	volume = {199},
	issn = {0008-8846},
	shorttitle = {Fracture in concrete},
	doi = {10.1016/j.cemconres.2025.108012},
	journal = {Cement and Concrete Research},
	author = {Mishra, A. and Carrara, P. and Griffa, M. and De Lorenzis, L.},
	year = {2026},
	pages = {108012},
}

@article{horne_enhanced_2025,
	title = {Enhanced geothermal systems for clean firm energy generation},
	volume = {1},
	copyright = {2025 Springer Nature Limited},
	issn = {3005-0685},
	doi = {10.1038/s44359-024-00019-9},
	number = {2},
	journal = {Nature Reviews Clean Technology},
	author = {Horne, Roland and Genter, Albert and McClure, Mark and Ellsworth, William and Norbeck, Jack and Schill, Eva},
	year = {2025},
	pages = {148--160},
}

@article{vafaie_chemo-hydro-mechanical_2023,
	title = {Chemo-hydro-mechanical effects of {CO2} injection on reservoir and seal rocks: {A} review on laboratory experiments},
	volume = {178},
	issn = {1364-0321},
	shorttitle = {Chemo-hydro-mechanical effects of {CO2} injection on reservoir and seal rocks},
	doi = {10.1016/j.rser.2023.113270},
	journal = {Renewable and Sustainable Energy Reviews},
	author = {Vafaie, Atefeh and Cama, Jordi and Soler, Josep M. and Kivi, Iman R. and Vilarrasa, Victor},
	year = {2023},
	pages = {113270},
}

@article{rutqvist_numerical_2005,
	series = {Research results from the {Decovalex} {III} \& {Benchpar} projects},
	title = {A numerical study of {THM} effects on the near-field safety of a hypothetical nuclear waste repository—{BMT1} of the {DECOVALEX} {III} project. {Part} 3: {Effects} of {THM} coupling in sparsely fractured rocks},
	volume = {42},
	issn = {1365-1609},
	doi = {10.1016/j.ijrmms.2005.03.012},
	number = {5},
	journal = {International Journal of Rock Mechanics and Mining Sciences},
	author = {Rutqvist, J. and Chijimatsu, M. and Jing, L. and Millard, A. and Nguyen, T. S. and Rejeb, A. and Sugita, Y. and Tsang, C. F.},
	year = {2005},
	pages = {745--755},
}

@article{wu_crack_2023,
	title = {Crack nucleation and propagation of electromagneto-thermo-mechanical fracture in bulk superconductors during magnetization},
	volume = {172},
	issn = {0022-5096},
	doi = {10.1016/j.jmps.2022.105168},
	journal = {Journal of the Mechanics and Physics of Solids},
	author = {Wu, Jian-Ying and Hong, Yi-Feng},
	year = {2023},
	pages = {105168},
}

@article{ruan_thermo-mechanical_2023,
	title = {A thermo-mechanical phase-field fracture model: {Application} to hot cracking simulations in additive manufacturing},
	volume = {172},
	issn = {0022-5096},
	shorttitle = {A thermo-mechanical phase-field fracture model},
	doi = {10.1016/j.jmps.2022.105169},
	journal = {Journal of the Mechanics and Physics of Solids},
	author = {Ruan, Hui and Rezaei, Shahed and Yang, Yangyiwei and Gross, Dietmar and Xu, Bai-Xiang},
	year = {2023},
	pages = {105169},
}

@incollection{ilgen_coupled_2019,
	title = {Coupled {Chemical}-{Mechanical} {Processes} {Associated} {With} the {Injection} of {CO2} into {Subsurface}},
	isbn = {978-0-12-812752-0},
	booktitle = {Science of {Carbon} {Storage} in {Deep} {Saline} {Formations}},
	publisher = {Elsevier},
	author = {Ilgen, Anastasia G. and Newell, Pania and Hueckel, Tomasz and Espinoza, D. Nicolas and Hu, Manman},
	editor = {Newell, Pania and Ilgen, Anastasia G.},
	year = {2019},
	doi = {10.1016/B978-0-12-812752-0.00015-0},
	pages = {337--359},
}

@article{ciantia_weathering_2013,
	title = {Weathering of submerged stressed calcarenites: chemo-mechanical coupling mechanisms},
	volume = {63},
	issn = {0016-8505},
	shorttitle = {Weathering of submerged stressed calcarenites},
	doi = {10.1680/geot.SIP13.P.024},
	number = {9},
	journal = {Géotechnique},
	author = {Ciantia, Matteo Oryem and Hueckel, Tomasz},
	year = {2013},
	pages = {768--785},
}

@article{stefanou_chemically_2014,
	title = {Chemically induced compaction bands: {Triggering} conditions and band thickness},
	volume = {119},
	copyright = {©2014. American Geophysical Union. All Rights Reserved.},
	issn = {2169-9356},
	shorttitle = {Chemically induced compaction bands},
	doi = {10.1002/2013JB010342},
	number = {2},
	journal = {Journal of Geophysical Research: Solid Earth},
	author = {Stefanou, Ioannis and Sulem, Jean},
	year = {2014},
	pages = {880--899},
}

@article{miehe_phase_2016,
  title={A phase-field model for chemo-mechanical induced fracture in lithium-ion battery electrode particles},
  author={Miehe, Christian and Dal, H{\"u}sn{\"u} and Sch{\"a}nzel, L-M and Raina, A},
  journal={International Journal for Numerical Methods in Engineering},
  volume={106},
  number={9},
  pages={683--711},
  year={2016},
  publisher={Wiley Online Library},
  doi = {10.1002/nme.5133},
}

@book{economides_reservoir_2000,
  title={Reservoir stimulation},
  author={Economides, Michael J and Nolte, Kenneth G and others},
  volume={18},
  year={2000},
  publisher={Wiley New York}
}

@article{castellanza_oedometric_2004,
	title = {Oedometric {Tests} on {Artificially} {Weathered} {Carbonatic} {Soft} {Rocks}},
	volume = {130},
	copyright = {Copyright © 2004 American Society of Civil Engineers},
	issn = {1090-0241},
	doi = {10.1061/(ASCE)1090-0241(2004)130:7(728)},
	number = {7},
	journal = {Journal of Geotechnical and Geoenvironmental Engineering},
	author = {Castellanza, Riccardo and Nova, Roberto},
	year = {2004},
	pages = {728--739},
}

@article{ciantia_effects_2015,
	title = {Effects of mineral suspension and dissolution on strength and compressibility of soft carbonate rocks},
	volume = {184},
	issn = {0013-7952},
	doi = {10.1016/j.enggeo.2014.10.024},
	journal = {Engineering Geology},
	author = {Ciantia, Matteo Oryem and Castellanza, Riccardo and Crosta, Giovanni B. and Hueckel, Tomasz},
	year = {2015},
	pages = {1--18},
}

@article{li_meso-scale_2024,
	title = {Meso-scale size effects of material heterogeneities on crack propagation in brittle solids: {Perspectives} from phase-field simulations},
	volume = {296},
	issn = {0020-7683},
	doi = {10.1016/j.ijsolstr.2024.112795},
	journal = {International Journal of Solids and Structures},
	author = {Li, Liuchi and Rao, Jack and Hufnagel, Todd C. and Ramesh, K. T.},
	year = {2024},
	pages = {112795},
}

@article{s_fracture_2025,
	title = {Fracture behavior and toughening mechanisms of brittle composites with stiff and compliant inclusions: {A} dynamic phase field study},
	volume = {321},
	issn = {0020-7683},
	doi = {10.1016/j.ijsolstr.2025.113545},
	journal = {International Journal of Solids and Structures},
	author = {Anupama, S. and Kishen, J. M. Chandra},
	year = {2025},
	pages = {113545},
}

@article{zietlow_measurement_1998,
	title = {Measurement of the intrinsic process zone in rock using acoustic emission},
	volume = {35},
	issn = {1365-1609},
	doi = {10.1016/S0148-9062(97)00323-9},
	number = {3},
	journal = {International Journal of Rock Mechanics and Mining Sciences},
	author = {Zietlow, W. K and Labuz, J. F},
	year = {1998},
	pages = {291--299},
}

@article{backers_tensile_2005,
	series = {Rock {Physics} and {Geomechanics}},
	title = {Tensile fracture propagation and acoustic emission activity in sandstone: {The} effect of loading rate},
	volume = {42},
	issn = {1365-1609},
	shorttitle = {Tensile fracture propagation and acoustic emission activity in sandstone},
	doi = {10.1016/j.ijrmms.2005.05.011},
	number = {7},
	journal = {International Journal of Rock Mechanics and Mining Sciences},
	author = {Backers, T. and Stanchits, S. and Dresen, G.},
	year = {2005},
	pages = {1094--1101},
}

@article{chen_single_2024,
	title = {Single fluid-driven crack propagation in analogue rock assisted by chemical environment},
	volume = {37},
	issn = {2352-3808},
	doi = {10.1016/j.gete.2023.100526},
	journal = {Geomechanics for Energy and the Environment},
	author = {Chen, Jing and Hu, Manman},
	year = {2024},
	pages = {100526},
}

@inproceedings{chen_chemically_2025,
  TITLE = {{Chemically enhanced fluid-driven crack propagation in analogue rock}},
  AUTHOR = {Chen, Jing and Hu, Manman},
  URL = {https://hal.science/hal-05343935},
  BOOKTITLE = {{3rd International Conference on Energy Geotechnics 2025}},
  ADDRESS = {Paris, France},
  ORGANIZATION = {{Navier Laboratory, Ecole Nationale des Ponts et Chauss{\'e}es, Universit{\'e} Gustave Eiffel, CNRS}},
  HAL_LOCAL_REFERENCE = {102755},
  YEAR = {2025},
  HAL_ID = {hal-05343935},
  HAL_VERSION = {v1},
}

@article{kristensen_phase_2020,
	title = {A phase field model for elastic-gradient-plastic solids undergoing hydrogen embrittlement},
	volume = {143},
	issn = {0022-5096},
	doi = {10.1016/j.jmps.2020.104093},
	journal = {Journal of the Mechanics and Physics of Solids},
	author = {Kristensen, Philip K. and Niordson, Christian F. and Martínez-Pañeda, Emilio},
	year = {2020},
	pages = {104093},
}

@article{miehe_phase_2016-1,
	title = {Phase field modeling of fracture in multi-physics problems. {Part} {III}. {Crack} driving forces in hydro-poro-elasticity and hydraulic fracturing of fluid-saturated porous media},
	volume = {304},
	issn = {0045-7825},
	doi = {10.1016/j.cma.2015.09.021},
	journal = {Computer Methods in Applied Mechanics and Engineering},
	author = {Miehe, Christian and Mauthe, Steffen},
	year = {2016},
	pages = {619--655},
}

@article{borden_phase-field_2016,
	series = {Phase {Field} {Approaches} to {Fracture}},
	title = {A phase-field formulation for fracture in ductile materials: {Finite} deformation balance law derivation, plastic degradation, and stress triaxiality effects},
	volume = {312},
	issn = {0045-7825},
	doi = {10.1016/j.cma.2016.09.005},
	journal = {Computer Methods in Applied Mechanics and Engineering},
	author = {Borden, Michael J. and Hughes, Thomas J. R. and Landis, Chad M. and Anvari, Amin and Lee, Isaac J.},
	year = {2016},
	pages = {130--166},
}

@article{bourdin_variational_2008,
	title = {The {Variational} {Approach} to {Fracture}},
	volume = {91},
	issn = {1573-2681},
	doi = {10.1007/s10659-007-9107-3},
	number = {1},
	journal = {Journal of Elasticity},
	author = {Bourdin, Blaise and Francfort, Gilles A. and Marigo, Jean-Jacques},
	year = {2008},
	pages = {5--148},
}

@article{liu2018,
    title={A hybrid finite volume and extended finite element method for hydraulic fracturing with cohesive crack propagation in quasi-brittle materials},
    author={Liu, Chong and Shen, Zhenzhong and Gan, Lei and Jin, Tian and Zhang, Hongwei and Liu, Detan},
    journal={Materials},
    volume={11},
    number={10},
    pages={1921},
    year={2018},
	doi = {10.3390/ma11101921}
}

@article{ambrosio_approximation_1990,
	title = {Approximation of functional depending on jumps by elliptic functional via t-convergence},
	volume = {43},
	copyright = {Copyright © 1990 Wiley Periodicals, Inc., A Wiley Company},
	issn = {1097-0312},
	doi = {10.1002/cpa.3160430805},
	number = {8},
	journal = {Communications on Pure and Applied Mathematics},
	author = {Ambrosio, Luigi and Tortorelli, Vincenzo Maria},
	year = {1990},
	pages = {999--1036},
}

@article{amor_regularized_2009,
	title = {Regularized formulation of the variational brittle fracture with unilateral contact: {Numerical} experiments},
	volume = {57},
	issn = {0022-5096},
	doi = {10.1016/j.jmps.2009.04.011},
	number = {8},
	journal = {Journal of the Mechanics and Physics of Solids},
	author = {Amor, Hanen and Marigo, Jean-Jacques and Maurini, Corrado},
	year = {2009},
	pages = {1209--1229},
}

@article{liu_generalized_2026,
	title = {A generalized higher-order phase-field model for brittle fracture in anisotropic rocks},
	volume = {210},
	issn = {0022-5096},
	doi = {10.1016/j.jmps.2026.106526},
	journal = {Journal of the Mechanics and Physics of Solids},
	author = {Liu, Sijia and Wang, Yunteng and Geng, Xueyu and Wu, Wei},
	year = {2026},
	pages = {106526},
}

@article{kristensen_assessment_2021,
	title = {An assessment of phase field fracture: crack initiation and growth},
	volume = {379},
	doi = {10.1098/rsta.2021.0021},
	number = {2203},
	journal = {Philosophical Transactions of the Royal Society A: Mathematical, Physical and Engineering Sciences},
	author = {Kristensen, Philip K. and Niordson, Christian F. and Martínez-Pañeda, Emilio},
	year = {2021},
	pages = {20210021},
}

@article{wu_influence_2025,
	title = {Influence of {Carbon} {Dioxide} on {Micro}-{Cracking} in {Calcite}: {An} {Atomistic} {Scale} {Investigation}},
	volume = {130},
	copyright = {© 2025 The Author(s).},
	issn = {2169-9356},
	shorttitle = {Influence of {Carbon} {Dioxide} on {Micro}-{Cracking} in {Calcite}},
	doi = {10.1029/2024JB030896},
	number = {4},
	journal = {Journal of Geophysical Research: Solid Earth},
	author = {Wu, Fanyu and Hu, Manman},
	year = {2025},
	pages = {e2024JB030896},
}

@article{hueckel_feedback_2009,
	title = {Feedback mechanisms in chemo-mechanical multi-scale modeling of soil and sediment compaction},
	volume = {36},
	issn = {0266-352X},
	doi = {10.1016/j.compgeo.2009.02.005},
	number = {6},
	journal = {Computers and Geotechnics},
	author = {Hueckel, Tomasz and Hu, Liang Bo},
	year = {2009},
	pages = {934--943},
}

@article{korec_phase-field_2023,
	title = {A phase-field chemo-mechanical model for corrosion-induced cracking in reinforced concrete},
	volume = {393},
	issn = {0950-0618},
	doi = {10.1016/j.conbuildmat.2023.131964},
	journal = {Construction and Building Materials},
	author = {Korec, Evžen and Jirásek, Milan and Wong, Hong S. and Martínez-Pañeda, Emilio},
	year = {2023},
	pages = {131964},
}

@article{xu_time-dependent_2025,
	title = {Time-{Dependent} {Deformation} of {Sandstone} due to {Chemical} {Corrosion}: {An} {Investigation} {Using} the {Heterogeneous} {Grain}-{Based} {Phase}-{Field} {Method}},
	volume = {58},
	issn = {1434-453X},
	doi = {10.1007/s00603-025-04439-2},
	number = {6},
	journal = {Rock Mechanics and Rock Engineering},
	author = {Xu, Bin and Xu, Tao and Du, Shigui and Heap, Michael J. and Kushnir, Alexandra R. L. and Liu, Ben},
	year = {2025},
	pages = {7013--7039},
}

@article{borja_da_rocha_damage_2026,
	title = {Damage localization as a bifurcation phenomenon and the resulting fracture patterns in soft materials},
	volume = {212},
	issn = {0022-5096},
	doi = {10.1016/j.jmps.2026.106567},
	journal = {Journal of the Mechanics and Physics of Solids},
	author = {Borja Da Rocha, Hudson and Cohen, Tal},
	year = {2026},
	pages = {106567},
}

@article{tanne_crack_2018,
	title = {Crack nucleation in variational phase-field models of brittle fracture},
	volume = {110},
	issn = {0022-5096},
	doi = {10.1016/j.jmps.2017.09.006},
	journal = {Journal of the Mechanics and Physics of Solids},
	author = {Tanné, E. and Li, T. and Bourdin, B. and Marigo, J. -J. and Maurini, C.},
	year = {2018},
	pages = {80--99},
}

@article{pham_gradient_2011,
	title = {Gradient {Damage} {Models} and {Their} {Use} to {Approximate} {Brittle} {Fracture}},
	volume = {20},
	issn = {1056-7895},
	doi = {10.1177/1056789510386852},
	number = {4},
	journal = {International Journal of Damage Mechanics},
	publisher = {SAGE Publications Ltd STM},
	author = {Pham, Kim and Amor, Hanen and Marigo, Jean-Jacques and Maurini, Corrado},
	year = {2011},
	pages = {618--652},
}

@article{liu2026role,
  title={The role of G-annulus in the interaction of fluid-driven cracks in porous media},
  author={Liu, Chong and Chen, Jing and Regenauer-Lieb, Klaus and Hu, Manman},
  journal={Acta Geotechnica},
  volume = {21},
  pages={3183–3207},
  year={2026},
  doi = {10.1007/s11440-026-02997-5},
  publisher={Springer}
}

@article{chen_fluid_2026,
	title = {Fluid-mediated process zone coalescence drives repulsive cracks into mutual attraction},
	volume = {214},
	issn = {0022-5096},
	doi = {10.1016/j.jmps.2026.106676},
	journal = {Journal of the Mechanics and Physics of Solids},
	author = {Chen, Jing and Wu, Fanyu and Hu, Manman},
	year = {2026},
	pages = {106676},
}

@book{zoback_reservoir_2007,
	title = {Reservoir Geomechanics},
	isbn = {978-0-521-14619-7},
	publisher = {Cambridge University Press},
	author = {Zoback, Mark D.},
	year = {2007},
	doi = {10.1017/CBO9780511586477},
}

@article{chen2023shear,
  title={Shear instability in geothermal fields under hydromechanical feedback},
  author={Chen, Jing and Tang, Xiaojie and Hu, Manman},
  journal={Geophysics},
  volume={88},
  number={5},
  pages={WB151--WB161},
  year={2023},
  doi = {10.1190/geo2022-0630.1},
}

@article{buscarnera_chemo-mechanics_2016,
	title = {Chemo-mechanics of cemented granular solids subjected to precipitation and dissolution of mineral species},
	volume = {40},
	copyright = {Copyright © 2015 John Wiley \& Sons, Ltd.},
	issn = {1096-9853},
	doi = {10.1002/nag.2486},
	number = {9},
	journal = {International Journal for Numerical and Analytical Methods in Geomechanics},
	author = {Buscarnera, Giuseppe and Das, Arghya},
	year = {2016},
	pages = {1295--1320},
}

@article{wu_companion,
  title={Pattern formation of parallel cracks in carbonate rocks: {A} dissolution-triggered instability},
  author={Wu, Fanyu and Veveakis, Manolis and Hu, Manman},
  journal={submitted},
  year={2026},
  pubstate = {submitted}
}

\end{document}